\documentclass[twocolumn,aps,prb,nofootinbib,superscriptaddress]{revtex4-2}
\usepackage{amsmath}
\usepackage{amssymb}
\usepackage{amsthm}
\usepackage{setspace}
\usepackage{graphicx}
\usepackage{braket}
\usepackage{mathrsfs}
\usepackage{newtxmath}
\usepackage{float}
\usepackage[colorlinks = true,linkcolor = red,urlcolor  = blue,citecolor = blue,anchorcolor = blue]{hyperref}
\usepackage[utf8]{inputenc}
\usepackage[english]{babel}
\usepackage{bm}
\usepackage[caption=false]{subfig}
\usepackage{appendix}
\usepackage{natbib}

\graphicspath{ {Figures/} }

\begin{document}

\title{Planar Josephson junction devices with narrow superconducting strips: Topological properties and optimization} 

\author{Purna P. Paudel}
\affiliation{Department of Physics and Astronomy, West Virginia University, Morgantown, WV 26506, USA}
\author{Javad Shabani}
\affiliation{Center for Quantum Information Physics, Department of Physics, New York University, New York, NY 10003, USA}
\author{Tudor D. Stanescu}
\affiliation{Department of Physics and Astronomy, West Virginia University, Morgantown, WV 26506, USA}

\begin{abstract}
We study the low-energy physics of planar Josephson junction structures realized in a quasi-two dimensional semiconductor system proximity-coupled to narrow superconducting films. Using both a recursive Green's function approach and an effective Hamiltonian approximation, we investigate the topological superconducting phase predicted to emerge in this type of system. We first characterize the effects associated with varying the electrostatic potentials applied within the unproximitized semiconductor regions. We then address the problem of optimizing the width of the superconductor films and identifying the optimal regimes characterized by large topological gap values. We find that structures with narrow superconducting films of widths ranging between about $100~$nm and $200~$nm can support topological superconducting phases with gaps up to $40\%$ of the parent superconducting gap, significantly larger than those characterizing the corresponding wide-superconductor structures. This work represents the first component of a proposed comprehensive strategy to address this optimization problem in  planar Josephson junction structures and realize robust topological devices.  
\end{abstract}

\maketitle

\section{Introduction} \label{Introduction}

Nature was rather stingy with providing robust, ready-to-use topological superconductors, but left open the door for engineering them using garden-variety materials, including technologically-friendly semiconductors proximity-coupled to ordinary s-wave superconductors \cite{Sau2010a,Lutchyn2010,Oreg2010,Sau2010,Alicea2010,Pientka2017,Hell2017}. This opportunity is extremely appealing, as the Majorana zero modes (MZMs) \cite{Read2000,Kitaev2001} hosted by topological superconductors---particle-hole symmetric excitations that obey non-Abelian statistics \cite{Leinaas1977,Wilczek1982} and represent a condensed matter analog of the elusive Majorana fermion \cite{Majorana1937}---provide a promising platform for fault-tolerant topological quantum computation \cite{Bravyi2002,Kitaev2003,Nayak2008,DSarma2015}.  

So far, the quest for topological superconductivity and MZMs in semiconductor-superconductor (SM-SC) hybrid nanowires 
\cite{Mourik2012, Deng2012, Das2012,Finck2013,Churchill2013,Deng2018,Chen2019,
Vaitiekenas2018,Grivnin2019,Yu2021,Microsoft2023} has generated remarkable progress in materials growth and device engineering and has provoked an acute awareness of the subtleties associated with correctly identifying signatures of nontrivial topology, but has not resulted in a clear demonstration of topologically-protected MZMs. The door toward engineering topological heterostructures proves to be rather narrow. This should not be surprising, considering that the relevant energy scales are in the range $\muup$eV--meV and, consequently, realizing the desired quantum states requires exquisite control of materials/interface quality and device engineering. The most significant challenge facing the demonstration of MZMs is overcoming the effects of inhomogeneity and disorder, which can destabilize the topological  phase hosting the MZMs and can generate topologically trivial states with signatures mimicking the presence of MZMs \cite{Motrunich2001,Brouwer2011,Kells2012,Liu2012,Pikulin2012,Rieder2013,Takei2013,DeGottardi2013,Adagideli2014,Stanescu2019,Ahn2021,Pan2022,Zeng2022,Roy2024}. 

In addition to improving the quality of the materials and the device engineering process, a critical task for realizing robust topological SM-SC structures is to identify optimal parameter regimes for building and operating the hybrid devices. For example, enhancing the gap that protects the topological phase may increase its robustness against disorder. Furthermore, operating in a regime that supports low-energy modes with characteristic length scales (much) shorter than the size of the system is necessary for topological protection. If these characteristic length scales are comparable to the system size, disorder can induce partially overlapping Majorana modes \cite{Zeng2022} with spectral weights (mostly) distributed within different halves of the system, which are capable of mimicking not only the local, but also some of the non-local signatures of MZMs \cite{DSarma2023}. Such a system is neither topological nor trivial, it is simply too short. An observation of Majorana-like signatures in such a system would be a premature reason for claiming victory, since extending the size of the system may lead to a (trivial) partially-separated Andreev bound state (ABS) \cite{Stanescu2019,Zeng2022}, instead of a much-desired pair of topologically-protected MZMs, if the underlying disorder is not compatible with a genuine topological phase. 

In this paper, we address the optimization problem in the context of planar SM-SC Josephson junction (JJ) structures \cite{Pientka2017,Hell2017,Stern2019,Setiawan2019a,Setiawan2019b,Scharf2019,Laeven2020,Paudel2021,Lesser2021, Oshima2022, Pekerten2022,Abboud2022,Luethi2023,Melo2023, Pekerten2024,Schiela2024}, which consist of a two-dimensional electron gas (2DEG) hosted by an SM quantum well proximity coupled to s-wave superconductors  
\cite{Shabani2016,Kjaergaard2016,Mayer2020}. These structures, proposed as a potential alternative to Majorana nanowires, have demonstrated promising evidence of key ingredients necessary for topological superconductivity \cite{Fornieri2019,Ren2019,Dartiailh2021,Banerjee2023}. Moreover, they have shown greater resilience against disorder compared to Majorana nanowires with similar planar design and comparable parameters \cite{Paudel2024}. This study, which is the first component of a more comprehensive optimization program (see Sec. \ref{Summ}), focuses on identifying the width of the SC films that maximizes the topological gap in clean JJ structures. We find that for systems with chemical potential values up to $\mu\sim 40~$meV, the optimal SC width is in the range $100-200~$nm, increasing weakly with $\mu$. The corresponding optimal values of the topological gap can be as high as $40\%$ of the parent SC gap, i.e., significantly larger than the topological gap values characterizing wide SC structures with similar materials and control parameters.  

The remainder of this paper is organized as follows. In Sec. \ref{Model} we describe our theoretical model and the methods used for obtaining the numerical solutions. The results are discussed in Sec. \ref{Results}, starting with a general description of the topological properties of a JJ structure with narrow SC films (Sec. \ref{General}) and continuing with an investigation of the main effects associated with varying the gate potentials applied within the unproximitized SM regions (Sec. \ref{SSB} and Sec. \ref{Crossover}). The main results of our optimization procedure are presented in Sec. \ref{TopoGap}. We conclude with a summary of our work and a discussion of future directions for the proposed optimization program (Sec. \ref{Summ}).

\section{Theoretical model} \label{Model}

\begin{figure}[t]
\begin{center}
\includegraphics[width=0.48\textwidth]{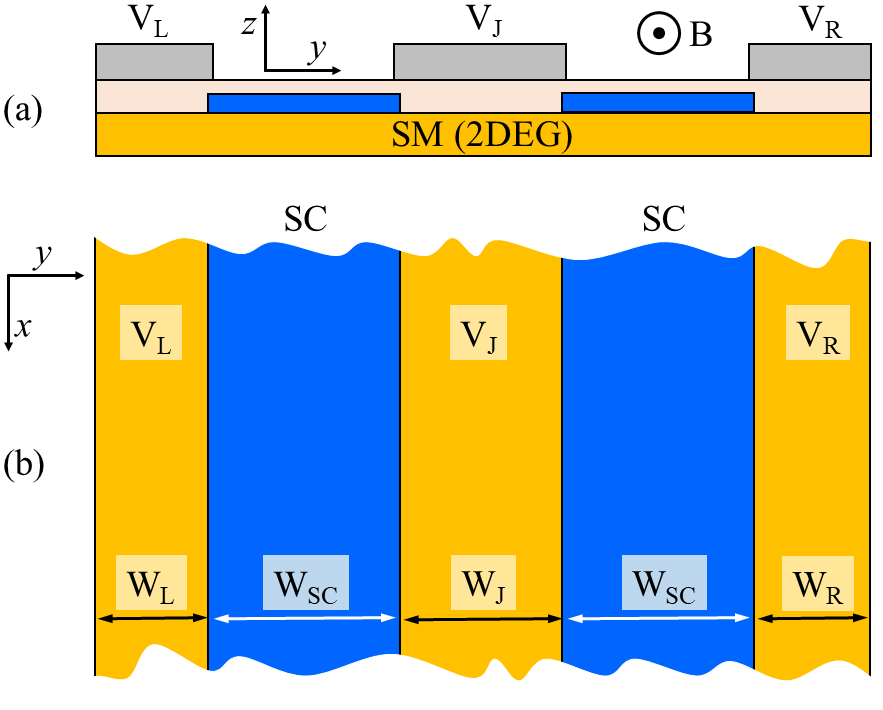}
\end{center}
\caption{Schematic representation of the planar SM-SC hybrid structure: (a) lateral view; (b) top view. A 2D electron gas hosted by a SM quantum well (orange) is proximity coupled to two thin SC films (blue) of width $W_{SC}$, forming an infinitely long Josephson junction of width $W_{J}$. Unproximitized SM regions (of widths $W_L$ and $W_R$) are present outside the SC films. The electrostatic potential in the regions not covered by the SCs is controlled by top gates (gray). An external magnetic field $B$ is applied in the $x-y$ plane (typically parallel to the junction, i.e., in the $x$ direction).}
\label{Fig1}
\vspace{0mm}
\end{figure}

In this paper, we study the low energy physics of semiconductor-superconductor (SM-SC) hybrid systems consisting of a planar semiconductor heterostructure that hosts a two-dimensional electron gas (2DEG) confined within a quantum well and proximity coupled to two narrow superconducting ribbons (see Fig. \ref{Fig1}). The resulting planar Josephson junction (JJ) is assumed to be infinitely long, the junction width is $W_J = 90~$nm, while the SC films are relatively narrow, $50 \leq W_{SC} \leq 400~$nm. The electrostatic potential within the regions not covered by the SC, i.e., the junction region and the left and right outside regions (see Fig. \ref{Fig1}), is controlled by top gates. Finally, a magnetic field is applied in the plane of the junction, typically in the $x$ direction (i.e., parallel to the junction), unless explicitly stated otherwise. 

The generic Hamiltonian describing the hybrid system has the form
\begin{equation}
H = H_{SM} + H_{SC} + H_{SM-SC}, \label{Htot}
\end{equation}
where the first two terms describe the semiconductor and superconductor components, respectively, and the last term characterizes the SM-SC coupling. The SM component, including the contributions from the applied external fields, is described by the (second quantized) nearest-neighbor tight-binding Hamiltonian 
\begin{eqnarray}
 H_{SM} &=& \sum_{k,j}\left(4t-2t\cos k a-\mu +V_j\right) \hat{c}_{k j}^\dagger \hat{c}_{k j}^{}  \label{HSM} \\
 &-&  t\!\!\sum_{k, \langle j, j^\prime\rangle}\hat{c}_{k j}^\dagger \hat{c}_{k j^\prime}^{} \!+\! \sum_{k,j}\left( \Gamma_x ~\!\hat{c}_{k j}^\dagger \sigma_x \hat{c}_{k j}^{} \!+\! \Gamma_y ~\!\hat{c}_{k j}^\dagger \sigma_y \hat{c}_{k j}^{}\right) \nonumber \\
 &+& \alpha \sum_{k,j}\left[\sin ka~\hat{c}_{k j}^\dagger \sigma_y \hat{c}_{k j}^{} \!+\! \frac{i}{2}\left(\hat{c}_{k j+1}^\dagger \sigma_x \hat{c}_{k j}^{} \!+ \!{\rm h.c.}\right)\right], \nonumber
\end{eqnarray}
where $j=1,2,\dots,N_y$ is the $y$-label of the site ${\bm r}=(i,j)$ of a square lattice with lattice constant $a$ that is translation invariant (infinitely long) along the $x$ direction, with $k\equiv k_x$ being the corresponding wave vector. Note that $\langle j,j^\prime\rangle$ designates nearest-neighbor sites. The electron creation and annihilation operators are $\hat{c}_{k j}^\dagger = (\hat{c}_{k j \uparrow}^\dagger, \hat{c}_{k j \downarrow}^\dagger)$ and $\hat{c}_{k j} = (\hat{c}_{k j \uparrow}, \hat{c}_{k j \downarrow})$, respectively, while $\sigma_x$ and $\sigma_y$ are Pauli matrices. The system parameters are the nearest-neighbor hopping ($t$), the chemical potential ($\mu$), and the Rashba spin-orbit coupling parameter ($\alpha$). Finally, the components of the (half) Zeeman splitting generated by the applied in-plane magnetic field ${\bm B}=(B_x, B_y)$ are $(\Gamma_x, \Gamma_y) = \frac{1}{2} g \mu_B (B_x, B_y)$, while the position-dependent electrostatic potential $V_j$ is zero in the SC regions and takes the values $V_L$, $V_J$, or $V_R$ in the unproximitized regions, as indicated in Fig. \ref{Fig1}. We note that the second quantized Hamiltonian (\ref{HSM}) can be written in terms of a first quantized $2N_y\times 2N_y$ Hamiltonian matrix $\widetilde{\cal H}_{SM}(k)$ as
\begin{equation}
H_{SM} = \sum_k \hat{\psi}_k^\dagger ~\!\widetilde{\cal H}_{SM}(k)~\! \hat{\psi}_k^{}, 
\end{equation}
with $\hat{\psi}_k = (\hat{c}_{k 1}^{}, \hat{c}_{k 2}^{}, \dots, \hat{c}_{k N_y}^{})^T$. Furthermore, $\widetilde{\cal H}_{SM}(k)$ can be ``expanded'' into a $4N_y\times 4N_y$ block-diagonal Bogoliubov--de Gennes matrix of the form ${\cal H}_{SM}(k) = \frac{1}{2}(\tau_z+\tau_0)\widetilde{\cal H}_{SM}(k) +\frac{1}{2}(\tau_z-\tau_0)\widetilde{\cal H}^*_{SM}(-k)$, where $\tau_z$ is a Pauli matrix associated with the particle-hole degree of freedom and $\tau_0$ is the $2\times 2$ unit matrix.

The coupling term $H_{SM-SC}$ in Eq. (\ref{HSM}) can be modeled as a tight binding Hamiltonian describing nearest-neighbor hopping (of amplitude $\tilde{t}$) across the SM-SC interface \cite{Paudel2024}, while $H_{SC}$, which corresponds to a mean-field description of the superconducting thin film, can be written as a Bogoliubov--de Gennes (BdG) tight-binding Hamiltonian on a cubic lattice, with nearest-neighbor hopping $t_{SC}$ and on-site pairing $\Delta_0~\! e^{\pm i\phi/2}$, where $\phi$ is the phase difference between the two superconductors \cite{Paudel2024}. It is convenient, however, to integrate out the SC degrees of freedom and describe the low-energy physics in terms of the effective Green's function $G_{SM}$ of the semiconductor. The SC contribution is incorporated as a self-energy \cite{Stanescu2022},
\begin{equation}
G_{SM}(\omega, k) = \left[\omega - {\cal H}_{SM}(k) -\Sigma_{SC}(\omega, k) \right]^{-1},    \label{Gwk}
\end{equation}
where ${\cal H}_{SM}(k)$ is the semiconductor BdG Hamiltonian matrix defined above and the unit matrix multiplying $\omega$ was omitted, for simplicity. The self-energy contribution, which is given by the Green's function of the SC film at the SM-SC interface (multiplied by $\tilde{t}^2$) \cite{Stanescu2022}, has the approximate form
\begin{equation}
\Sigma_{SC}(\omega) = -\frac{\gamma}{\sqrt{\Delta_0^2-\omega^2}}\left(\omega~\!\sigma_0\tau_0\mathbb{I}_{SC}^0 + \Delta_0~\!\sigma_y\tau_y \mathbb{I}_{SC}^\phi\right), \label{SigSC}
\end{equation}
where, as mentioned above, $\sigma_\mu$ and $\tau_\nu$ are Pauli matrices associated with the spin and particle-hole degrees of freedom, respectively.  $\mathbb{I}_{SC}^0$ and $\mathbb{I}_{SC}^\phi$ are $N_y\times N_y$ diagonal matrices with nonzero elements $[\mathbb{I}_{SC}^0]_{jj}=1$ and $[\mathbb{I}_{SC}^\phi]_{jj}=e^{\pm i\phi/2}$ if $j$ labels a site within the superconducting regions. We note that a robust proximity effect induced by a thin SC film requires the presence of disorder in the superconductor, e.g., the presence of surface roughness, and, consequently, the effective coupling $\gamma$ is a position-dependent, quasi-local matrix, $\gamma_{{\bm r} {\bm r}^\prime} \approx \gamma({\bm r})~\! \delta_{{\bm r} {\bm r}^\prime}$ \cite{Stanescu2022}. In this work, we neglect this position dependence (and the corresponding proximity-induced disorder) \cite{Stanescu2022} and consider the average effective coupling $\gamma = \langle \gamma({\bm r}) \rangle_{\bm r}$. 

The low-energy physics of the hybrid system is characterized by calculating numerically the total density of states (DOS), $\rho(\omega)$, and the local density of states (LDOS), $\rho(\omega, j)$, which can be expressed in terms of the spectral function at position $j$ across the structure,
\begin{equation}
A(\omega, k; j) = -\frac{1}{\pi}{\rm Im}~\!{\rm Tr}[G_{SM}(\omega+i\eta, k)]_{jj}, \label{Awk}
\end{equation}
as
\begin{equation}
 \rho(\omega)=\sum_{k, j} A(\omega, k; j), ~~~~~~~ \rho(\omega, j)=\sum_{k} A(\omega, k; j).
\end{equation}
Note that the trace in Eq. (\ref{Awk}) is over the spin and particle-hole labels, while the small positive parameter $\eta$ (typically on the order of $1~\muup$eV) introduces a finite broadening of the spectral features. To make the numerical evaluation of the Green's function more efficient, we implement the recursive Green's function approach described in Refs. \onlinecite{MacKinnon1985, Lewenkopf2013} for an effectively one-dimensional system of length $L_y=N_y a$ corresponding to each relevant $k$ value in Eq. (\ref{Gwk}), i.e., for $|k|\lesssim k_F$, where $k_F$ is the Fermi wave vector of the semiconductor. 

In addition to the Green's function approach described above, we also characterize the system using an effective low-energy Hamiltonian obtained within the ``static approximation'' corresponding to $\sqrt{\Delta_0^2-\omega^2}\approx \Delta_0$ in Eq. (\ref{SigSC}). Explicitly, a spin and particle-hole block of the effective Hamiltonian corresponding to lattice indices $j$ and $j^\prime$ has the form
\begin{equation}
\left[{\cal H}_{eff}(k)\right]_{j j^\prime} \!\!=\!\!\left\{
\begin{array}{l}
~~~\!\left[{\cal H}_{SM}\right]_{j j^\prime}; ~~~~~~~~~~~~~~~~~~~~~ j,j^\prime \in\!SM, \\
\!Z^{}\left[{\cal H}_{SM}\right]_{j j^\prime}\!-\!\Delta~\!\sigma_y\tau_y\delta_{j j^\prime}; ~~~~~~ j,j^\prime \in\!SC, \\
\!Z^\frac{1}{2}\left[{\cal H}_{SM}\right]_{j j^\prime}; ~~~ j(j^\prime)\in \!SM, ~j^\prime(j)\in\!SC, \label{Heff}
\end{array}\right.
\end{equation}
where $\Delta = \gamma \Delta_0/(\gamma + \Delta_0)e^{\pm i\phi/2}$ is the induced pairing potential within the proximitized ($SC$) regions, $Z=\Delta_0/(\Delta_0+\gamma)$ is the quasiparticle residue, which corresponds to the weight of a low-energy state within the semiconductor component of the heterostructure \cite{Stanescu2017a}, while $j\in SM$ and $j \in SC$ designate sites within the unproximitized ($SM$) and proximitized ($SC$) regions, respectively. Note that all parameters (i.e., hopping, Rashba coefficient, Zeeman field, etc.) within the $SC$ regions are renormalized by a factor $Z$, while the nearest-neighbor contributions that couple the $SC$ and $SM$ regions (i.e., hopping and Rashba spin-orbit coupling along the $y$ direction) are renormalized by a factor $Z^\frac{1}{2}$. In the strong coupling regime, $\gamma > \Delta_0$, this proximity-induced low-energy renormalization can be significant. The low-energy spectrum, $E_n(k)$, obtained by diagonalizing the effective Hamiltonian is expected to be accurate for energies much smaller than the parent SC gap, $\Delta_0$, when the static approximation holds, while the parameters corresponding to zero-energy states can be obtained exactly. Thus, the effective Hamiltonian approach provides an alternative way of determining the topological phase diagram (which involves zero-energy states) and a convenient (approximate) method of calculating the size of the quasiparticle gap. The quantitative agreement between the two methods is excellent for energies up to about $100~\muup$eV (see below).

The numerical parameters used in the calculation are the following: the lattice constant of the square lattice is $a=5~$nm; the hopping parameter is $t=50.8~$meV, which corresponds to an effective mass $m^*=0.03m_0$, with $m_0$ being the free electron mass; the Rashba spin-orbit coefficient is $\alpha =5~$meV, corresponding to $\alpha\cdot a =250~$meV$\cdot$\AA; the parent SC gap is $\Delta_0=0.25~$meV; the effective SM-SC coupling is $\gamma=0.75~$meV, which corresponds to $\gamma=3\Delta_0$, i.e., strong coupling; the width of the junction region is fixed, $W_J=90~$nm, but we consider different widths of the SC regions, as specified in the next section; the chemical potential ($\mu$), Zeeman field ($\Gamma_x, \Gamma_y$), and applied gate potentials ($V_J, V_L, V_R$) are used as control parameters. The model parameters are consistent with the most developed epitaxial SM-SC system, which is based on InAs quantum wells proximitized with epitaxially grown Al \cite{Shabani2016}. We note that there is strong interest in broadening the scope of SM-SC hybrids, e.g., by using larger gap superconductors, such as Sn, Pb, and Nb compounds \cite{Pendharkar2021,Kanne2021,Drachmann2017}, or establishing hyperdoped Ga-Ge as a low-disorder, epitaxial superconductor-semiconductor platform \cite{Steele2024}, but significant challenges persist (e.g., ensuring a large enough value of the g-factor that would make the topological phase practically accessible). Should any of these new platforms become functional, the optimization procedure presented in Sec. IIID has to be repeated using the corresponding materials parameters.

\section{Numerical results}  \label{Results}

In this section, we discuss the results of our numerical analysis based on a combination of the Green's function method and the effective Hamiltonian approach described above. We start by comparing the two methods and discussing some generic features of the topological phase diagram (Sec. \ref{General}). The effect of applying a gate potential in the outside SM regions is discussed in Sec. \ref{SSB}, while the effect of depleting the junction region, which determines a crossover to a Majorana wire regime, is presented in Sec. \ref{Crossover}. Finally, in Sec. \ref{TopoGap}, we investigate how the width of the SC films affects the topological phase diagram and the topological gap and we identify the optimal regimes for maximizing the gap. 

\subsection{General properties} \label{General}

\begin{figure}[t]
\begin{center}
\includegraphics[width=0.48\textwidth]{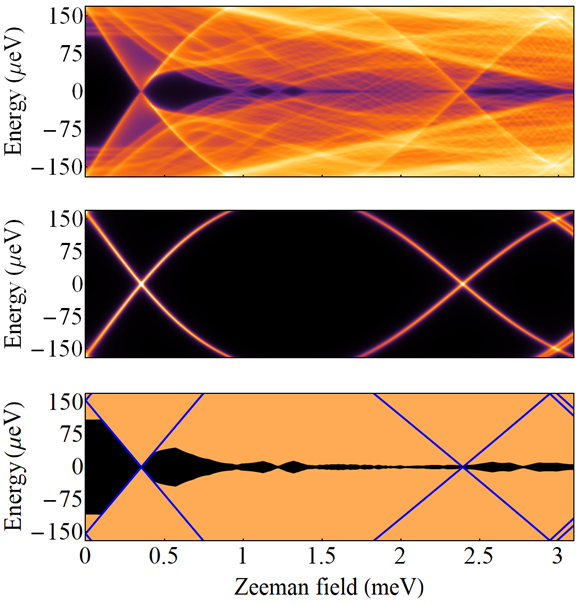}
\end{center}
\caption{{\em Top}: Density of states as a function of Zeeman field (applied parallel to the junction) and energy for a JJ structure with chemical potential $\mu=31.4~$meV and junction potential $V_J=25~$meV. The SC width is $W_{SC}=150~$nm and the outside SM regions are depleted, $V_L=V_R=100~$meV. The vanishing of the quasiparticle gap (at $\Gamma_x\approx 0.35~$meV and $\Gamma_x\approx 2.4~$meV) is associated with topological quantum phase transitions. For $\Gamma_x\gtrsim 1~$meV the topological gap is nearly zero (on the order of a few $\muup$eV). {\em Middle}: Zeeman field dependence of the spectral function at $k=0$. {\em Bottom}: Quasiparticle gap (black) and $k=0$ modes (blue lines) calculated using the effective Hamiltonian corresponding to the static approximation (see main text).}
\label{Fig2}
\vspace{-2mm}
\end{figure}

To compare the results obtained using the Green's function and the effective Hamiltonian methods and to justify using them interchangeably, as convenient, we consider the dependence of the low-energy spectrum on the applied Zeeman field for a JJ structure with chemical potential $\mu=31.4~$meV and phase difference $\phi=0$. The results are shown in Fig. \ref{Fig2} and the values of the geometric parameters and applied gate potentials are provided in the caption. The top panel, which shows the Zeeman field and energy dependence of the DOS, reveals a finite quasiparticle gap (of about $110~\muup$eV) in the absence of an applied Zeeman field. The gap is reduced in the presence of an applied field, closing and reopening at $\Gamma\approx 0.35~$meV, then closing and reopening again at $\Gamma\approx 2.4~$meV. The closing and reopening of the quasiparticle gap signals a topological quantum phase transition (between trivial and topological SC phases). Note, however, that a topological quantum phase transition (TQPT) is specifically associated with the closing and reopening of the gap at $k=0$; the closing (and reopening) of the gap at finite wave vectors is {\em not} associated with a TQPT. The Zeeman field and energy dependence of the spectral function at $k=0$ is shown in the middle panel of Fig. \ref{Fig2}, with the x-like zero energy crossings clearly marking the TQPTs. 

Indeed, in general the JJ structure is a quasi-one-dimensional system in symmetry class D, which has a $\mathbb{Z}_2$ topological classification. The Hamiltonian has particle-hole (charge conjugation) symmetry, $U_C^{} {\cal H}^*(k)U_C^{-1} = -{\cal H}(-k)$, with $U_C^{}$ being a unitary matrix of the form $U_C^{} =\mathbb{I}~\!\sigma_0\tau_x$, where $\mathbb{I}$ is the $N_y\times N_y$ identity matrix (associated with the spatial degrees of freedom), $\sigma_0$ is the spin identity matrix, and $\tau_x$ is a Pauli matrix associated with the particle-hole degree of freedom. The $\mathbb{Z}_2$ invariant can be defined as \cite{Kitaev2001}
\begin{equation}
\nu= {\rm sign}\{{\rm Pf}[H_{eff}(0)U_C]\}~{\rm sign}\{{\rm Pf}[H_{eff}(\pi)U_C]\} =\pm 1,  \label{nu}
\end{equation}
with $\nu=+1$ and $\nu=-1$ corresponding to the trivial and topological phases, respectively. In Eq. (\ref{nu}) ${\rm Pf}[\dots]$ designates the Pfaffian and $H_{eff}(k)U_C$ is a $4N_y\times4N_y$ skew-symmetric matrix. Using the property ${\rm Pf}[A^2]={\rm Det}[A]$, and the fact that $|{\rm Det}[U_C]|=1$, a TQPT, which corresponds to the vanishing of one of the Pfaffians in Eq. (\ref{nu}), implies ${\rm Det}[H_{eff}(k)]=0$ at $k=0$ or $k=\pi$. However, the system is characterized by a large gap at $k=\pi$ over the entire (relevant) parameter space. This implies that the TQPT is associated with the vanishing of the quasiparticle gap at $k=0$, i.e., ${\rm Det}[H_{eff}(0)]=0$, as mentioned above. 

\begin{figure}[t]
\begin{center}
\includegraphics[width=0.48\textwidth]{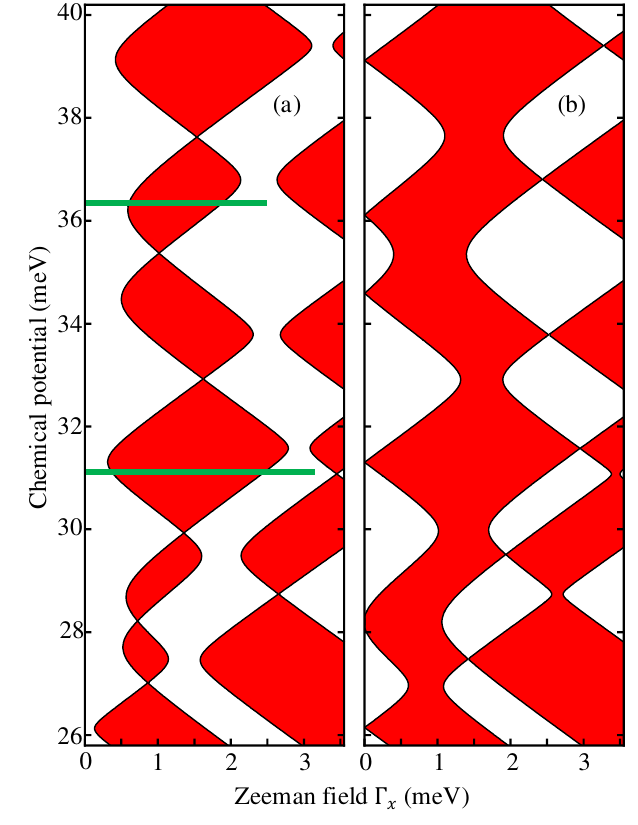}
\end{center}
\caption{Topological phase diagram as a function of Zeeman field ($\Gamma_x$) and chemical potential for a JJ device with $W_{SC}=150~$nm. The phase boundaries correspond to the vanishing of the (bulk) quasiparticle gap at $k=0$. The white areas are topologically trivial, while the red regions correspond to a topological superconducting phase. The gate potential on the junction is $V_J$  = 25 meV and the gate potentials outside the SC regions are $V_L = V_R = 100~$meV. The relative phase difference between the two narrow SCs is (a) $\phi=0$ and (b) $\phi=\pi$. The horizontal cut at $\mu=31.4~$meV (marked by a green line) corresponds to the low-energy spectra in Fig. \ref{Fig2}, while the DOS along the cut at $\mu=36.4~$meV is shown in Fig. \ref{Fig7}(a).}
\label{Fig3}
\vspace{-2mm}
\end{figure}

The lower panel in Fig. \ref{Fig2} shows the low-energy spectrum calculated using the effective Hamiltonian given by Eq. (\ref{Heff}). 
The $k=0$ modes (blue lines) are in excellent quantitative agreement with the $k=0$ spectral function shown in the middle panel for energies up to about $100~\muup$eV. In particular, the zero-energy crossings associated with the TQPT align precisely with those from the Green's function approach. Similarly, the quasiparticle gap, which corresponds to ${\rm min}_{n,k} [E_n(k)]$, is in excellent agreement with the results shown in the upper panel. Since the low-energy spectrum obtained using the effective Hamiltonian method does not depend on the broadening of spectral features (i.e., on the parameter $\eta$) and is not affected by the ``visibility'' of these features (i.e., their spectral weight within the SM component), we use the static approximation (and the corresponding effective Hamiltonian) to evaluate the size of the topological gap, for convenience.  

\begin{figure}[t]
\begin{center}
\includegraphics[width=0.48\textwidth]{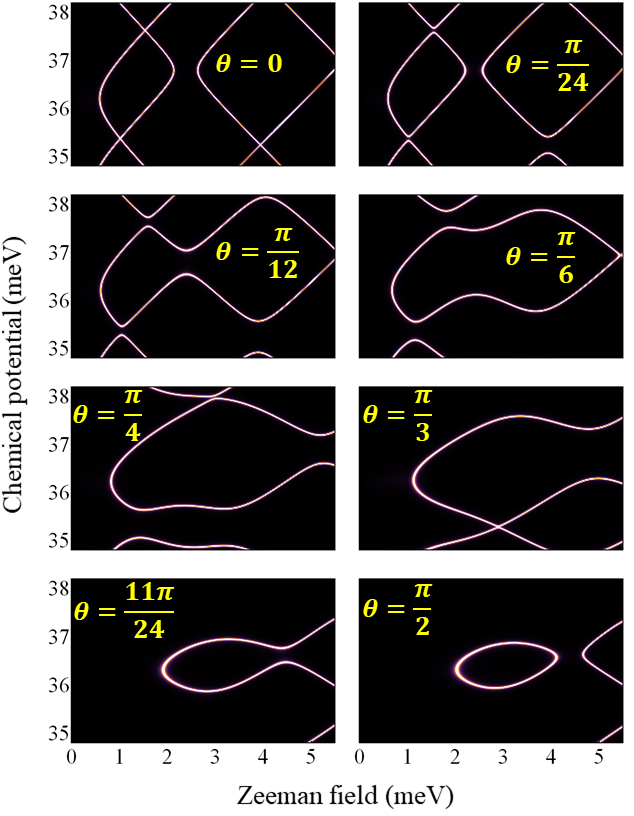}
\end{center}
\caption{Evolution of a section of the topological phase diagram in Fig. \ref{Fig3} as function of the angle $\theta$ between the applied Zeeman field and the direction parallel to the junction (i.e. the $x$-direction). The lines correspond to a finite spectral weight at zero energy and $k=0$, i.e., they indicate the vanishing of the quasiparticle gap at $k=0$. Note that within each ``loop'' the $\mathbb{Z}_2$ topological invariant $\nu$ given by Eq. (\ref{nu}) is nontrivial (i.e., $\nu=-1$, see Fig. \ref{Fig6}). However, the corresponding low-energy spectrum may be gapless (see Figs. \ref{Fig7} and \ref{Fig8}).}
\label{Fig4}
\vspace{-2mm}
\end{figure}

Next, we calculate the topological phase diagram for a system with SC width $W_{SC}=150~$nm and junction potential $V_J=25~$meV. Figure \ref{Fig3} shows the results for JJ devices with phase difference $\phi=0$ (left panel) and $\phi=\pi$ (right panel). The magnetic field is applied parallel to the junction, i.e., $\Gamma_y=0$. The area corresponding to the topological phase (red shading) is slightly larger for $\phi=\pi$ and extends to $\Gamma_x=0$ for certain specific values of the chemical potential.
We point out that, although the topological phase covers a significant percentage of the parameter space, the corresponding topological gaps are rather small, on the order of a few $\muup$eV and up to about $30~\muup$eV (see, e.g., Fig. \ref{Fig2}). Identifying the optimal parameters that maximize this gap is one of the main objectives of this work (see below, Sec. \ref{TopoGap}). 

\begin{figure}[t]
\begin{center}
\includegraphics[width=0.48\textwidth]{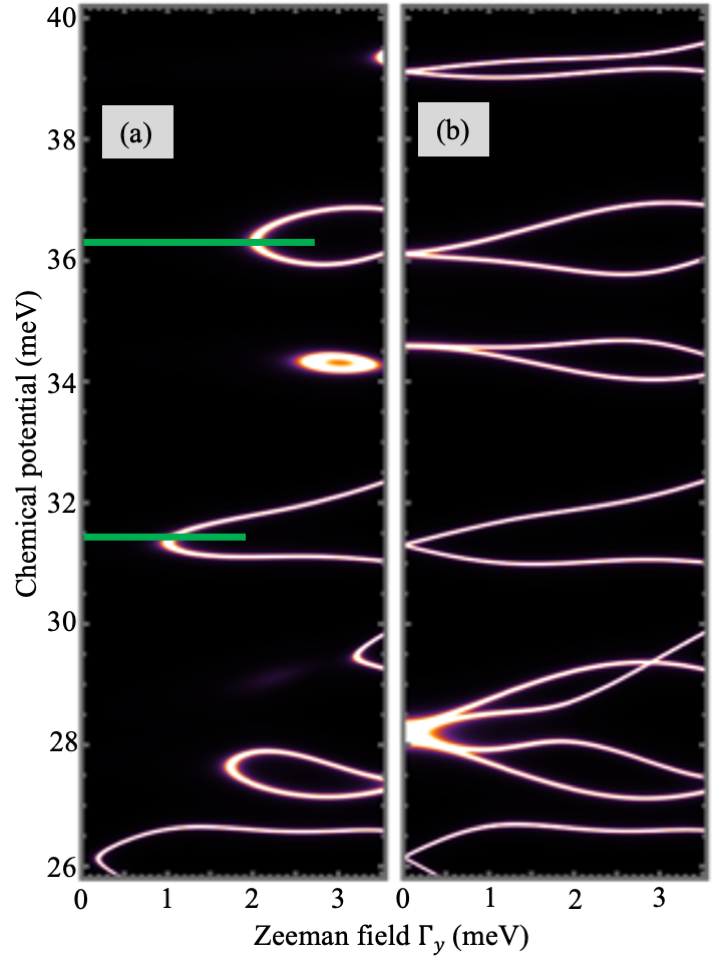}
\end{center}
\vspace{-2mm}
\caption{Phase diagram as a function of Zeeman field, $\Gamma_y$, applied perpendicular to the junction and chemical potential. The gate potential in the junction region is $V_J  = 25~$meV and the gate potential outside the SC regions is $V_L= V_R = 100~$meV. The relative phase difference between the two narrow SCs is (a) $\phi=0$  and (b) $\phi=\pi$. The light-colored lines correspond to a finite spectral weight at zero energy and $k=0$. The density of states along the horizontal cuts marked by green lines in (a), which correspond to $\mu = 36.4~$meV and $\mu=31.4~$meV, is shown in Fig. \ref{Fig7}, panels (b) and (c), respectively.}
\label{Fig5}
\vspace{-2mm}
\end{figure}

A natural question concerns the fate of the topological SC phase in the presence of an in-plane magnetic field applied along an arbitrary direction. While a detailed analysis is beyond the scope of this discussion, we highlight two key aspects concerning the topological phase boundaries and the quasiparticle gap. We consider an in-plane Zeeman field $\Gamma$ that makes an angle $\theta$ with the direction parallel to the junction, i.e., has components $\Gamma_x = \Gamma \cos \theta$  and $\Gamma_y = \Gamma \sin \theta$. 
Figure \ref{Fig4} shows the evolution of the phase boundary in the $\Gamma-\muup$ plane (for $34.8\leq \mu \leq 38.2~$meV and $\Gamma \leq 5.4~$meV) with the angle $\theta$. With increasing $\theta$, the phase boundaries, which correspond to a finite zero energy spectral weight at $k=0$ and typically form closed loops, change their topology. 
The boundaries corresponding to $\theta=\pi/2$,  i.e., Zeeman field applied perpendicular to the junction, are also shown in Fig. \ref{Fig5} for the same range of control parameters as in Fig. \ref{Fig3} and a superconducting phase difference $\phi=0$ [Fig. \ref{Fig5}(a)] and $\phi=\pi$ [Fig. \ref{Fig5}(b)]. Note that the phase boundary ``loops'' corresponding to $\phi=0$ are characterized by a finite minimum Zeeman energy, while those corresponding to $\phi=\pi/2$ can emerge from $\Gamma_y=0$. Also note that the ``loops'' in Fig. \ref{Fig5}(b) correspond to separated ``islands'' in the parameter space, while the low-field topological region in Fig. \ref{Fig3}(b), which corresponds to an orientation of the Zeeman field parallel to the junction, is path-connected (actually simply connected). 

\begin{figure}[t]
\begin{center}
\includegraphics[width=0.45\textwidth]{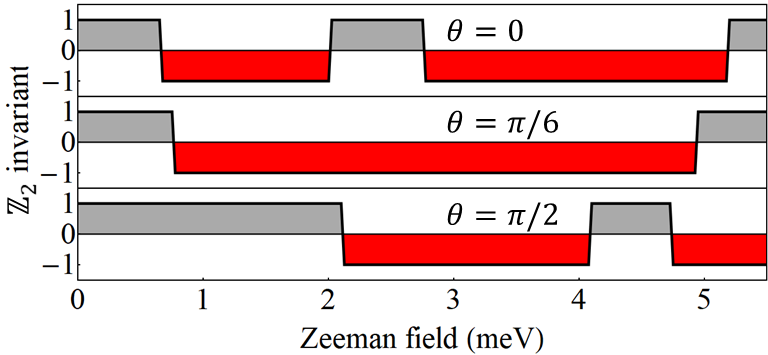}
\end{center}
\vspace{-2mm}
\caption{Topological invariant as a function of the applied Zeeman field along cuts (corresponding to $\mu=36.5~$meV) through representative diagrams in Fig. \ref{Fig4}. The topological invariant is $\nu=+1$ in the trivial phase and $\nu=-1$ in the topological phase (red shading).}
\label{Fig6}
\vspace{-3mm}
\end{figure}

To demonstrate that the features shown in Figs. \ref{Fig4} and  \ref{Fig5} are indeed phase boundaries, we calculate the \(\mathbb{Z}_2\) topological invariant given by Eq. (\ref{nu}) as a function of the applied Zeeman field along cuts (corresponding to $\mu = 36.5~$meV) through the diagrams with \(\theta = 0\), \(\pi/6\), and \(\pi/2\). The results shown in Fig. \ref{Fig6} clearly establish that, starting from a trivial phase (i.e., $\nu=+1$) at $\Gamma=0$, the invariant changes its sign each time one crosses a phase boundary. In other words, the phase inside the closed loops corresponding to $\theta=\pi/6$ and $\theta=\pi/2$ (see Fig. \ref{Fig4}) are topologically nontrivial ($\nu=-1$). Of course, for $\theta=0$ the topological regions are consistent with those marked by the red shading in Fig. \ref{Fig3}(a). 

\begin{figure}[t]
\begin{center}
\includegraphics[width=0.48\textwidth]{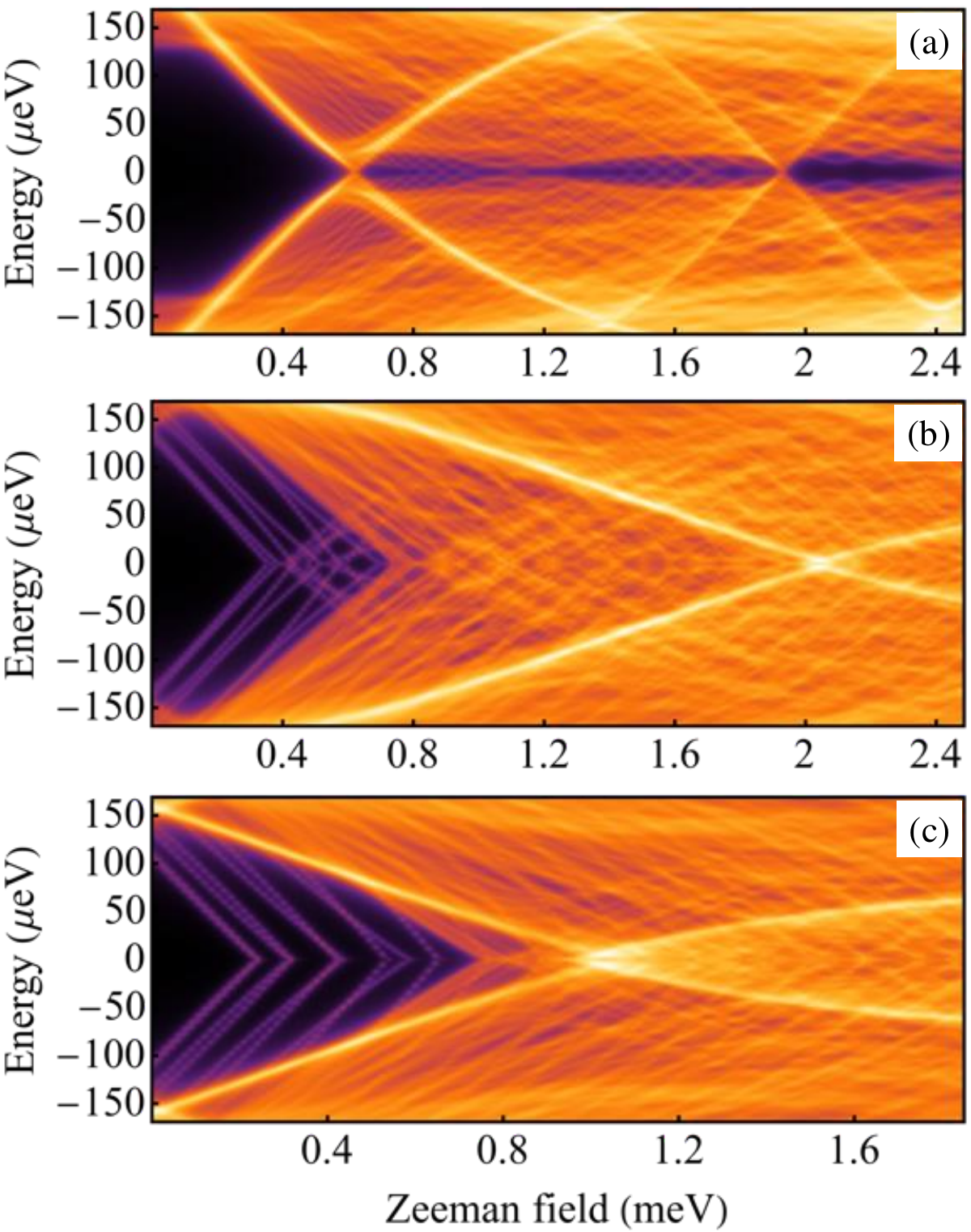}
\end{center}
\vspace{-3mm}
\caption{Density of states as a function of Zeeman field and energy calculated along the green cuts marked in  Fig. \ref{Fig3}  and  Fig. \ref{Fig6}.  (a) $\mu = 36.4 ~$meV and Zeeman field parallel to the junction ($\Gamma =\Gamma_x$); (b) $\mu =36.4~$meV, $\Gamma =\Gamma_y$; (c) $\mu = 31.4~$meV,  $\Gamma =\Gamma_y$. Note that $\Gamma =\Gamma_x$ results in a small (but finite) topological gap [also see Figs. \ref{Fig2} and \ref{Fig8}(a)], while for $\Gamma =\Gamma_y$, i.e., Zeeman field perpendicular to the junction, the system becomes gapless for $\Gamma_y \gtrsim 0.25~$meV [see Fig. \ref{Fig8}(b)].}
\label{Fig7}
\vspace{-2mm}
\end{figure}

A key difference between the nontrivial ($\nu=-1$) regions for $\theta=0$ (i.e., field parallel to the junction) and $\theta=\pi/2$ (i.e., field perpendicular to the junction) is that the former are typically gapped, while the later are gapless. To illustrate this important point, Fig. \ref{Fig7} presents the density of states (DOS) as a function of Zeeman field and energy calculated along representative cuts in Fig. \ref{Fig3} ($\theta=0$; $\mu=36.4~$meV) and Fig. \ref{Fig5}  ($\theta=\pi/2$; $\mu=36.4~$meV and $\mu=31.4~$meV). 
Note that the DOS corresponding to the cut with $\mu=31.4~$meV in Fig. \ref{Fig3} is shown in Fig. \ref{Fig2}. While the DOS corresponding to $\theta=0$ is characterized by small, but finite topological gaps [see Fig. \ref{Fig2} and Fig. \ref{Fig7}(a)], for $\theta=\pi/2$ the system becomes gapless [see Fig. \ref{Fig7}(b) and (c)].

\begin{figure}[t]
\begin{center}
\includegraphics[width=0.48\textwidth]{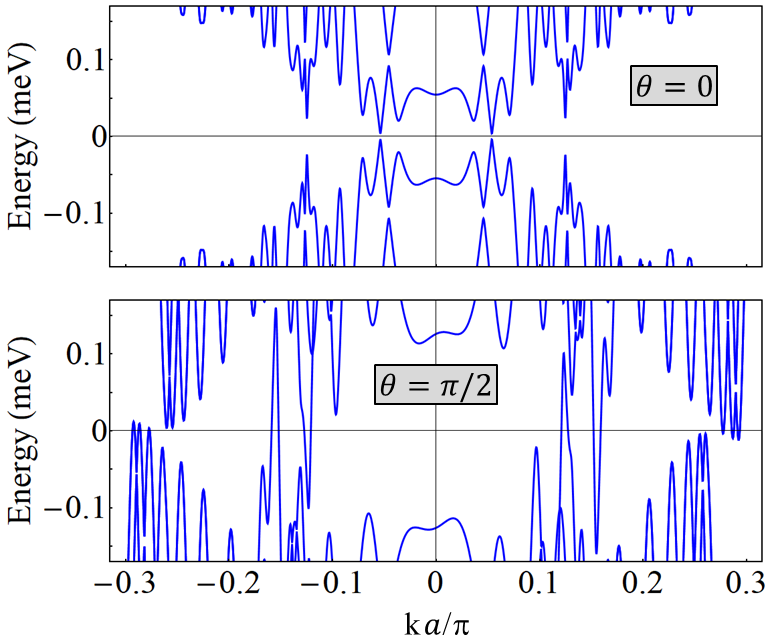}
\end{center}
\vspace{-2mm}
\caption{Low-energy spectrum as a function of $k$ for a system with $\mu=36.4~$meV and Zeeman field $\Gamma=0.8~$meV oriented (a) parallel and (b) perpendicular to the junction. Note that in the lower panel the gap closes at non-zero values of $k$.}
\label{Fig8}
\vspace{-2mm}
\end{figure}

To shed some light on this behavior, we calculate the low-energy spectrum as a function of the wave vector $k$ for two sets of parameters corresponding to $\Gamma_x=0.8~$meV in Fig. \ref{Fig7}(a) and $\Gamma_y=0.8~$meV in Fig. \ref{Fig7}(b), respectively. The results shown in Fig. \ref{Fig8} clearly illustrate the presence of a small quasiparticle gap for $\theta=0$ and the gapless nature of the $\theta=\pi/2$ state. More importantly, the spectra in Fig. \ref{Fig8} reveal that, while in general we have $E(k) = -E(-k)$, consistent with the presence of particle-hole symmetry, for $\theta=0$ the spectrum also satisfies the condition $E(k) = E(-k)$. This additional condition helps protecting the quasiparticle gaps at $k\neq 0$, although gapless states are still possible, particularly in systems with large chemical potential \cite{Paudel2024}. Note that for $\theta=0$ the gapless (or small gap) modes are typically associated with low $k$ values, $k a/\pi < 0.1$, i.e., with the top few occupied subbands (also see Appendix \ref{AppB}). By contrast, an arbitrary orientation of the Zeeman field ($\theta\neq 0$) favors the closing of the quasiparticle gap at large $k$ values (see Fig. \ref{Fig8}, bottom panel). Hence, the quasiparticle gap can close at relatively low values of the applied Zeeman field before the $k=0$ mode becomes gapless, i.e., without a topological quantum phase transition. The possibility of having small (or vanishing) quasiparticle gaps associated with finite-$k$ modes, which can impact the stability of the Majorana modes hosted by the structure, severely limits the parameter space for a robust topological superconducting phase and makes parameter optimization a critical step in the investigation of planar JJ structures (see below, Sec. \ref{TopoGap}). 

\subsection{Structures with undepleted outside SM regions} \label{SSB}

\begin{figure}[t]
\begin{center}
\includegraphics[width=0.48\textwidth]{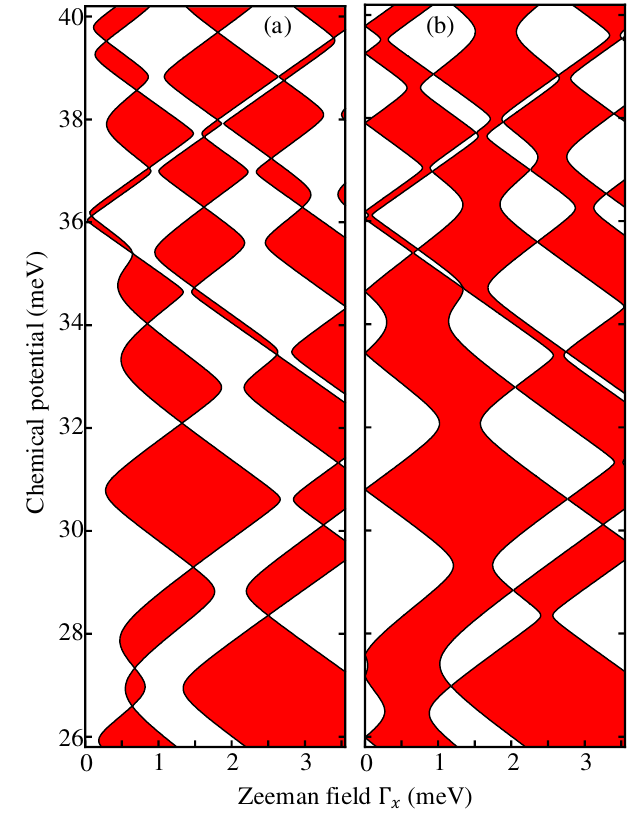}
\end{center}
\caption{Topological phase diagram as a function of Zeeman field ($\Gamma_x$) and chemical potential for a system with $W_{SC}=150~$nm,  $W_L= W_R= 100~$nm, and applied  gate potentials $V_J = 25~$meV and  $V_L= V_R= 35~$meV. The relative phase difference between the SC films is  (a) $\phi=0$ and (b) $\phi=\pi$. Note that for $\mu\gtrsim V_{L(R)}$ the electrons can access the outside SM regions, which results in a ``fragmentation'' of the topological phase diagram. This effect becomes stronger when widening the outside SM regions, i.e., increasing $W_L$ and $W_R$ (see Fig. \ref{Fig10}).}
\label{Fig9}
\vspace{-2mm}
\end{figure}

After the general discussion in the previous section, we focus on the effect of varying the applied electrostatic potential in the (unproximitized) SM regions, starting with the outside (left and right) regions (see Fig. \ref{Fig1}). For simplicity, we assume that the two regions have the same width, i.e., $W_L=W_R$, and the same potential value, $V_L=V_R$. 

\begin{figure}[t]
\begin{center}
\includegraphics[width=0.48\textwidth]{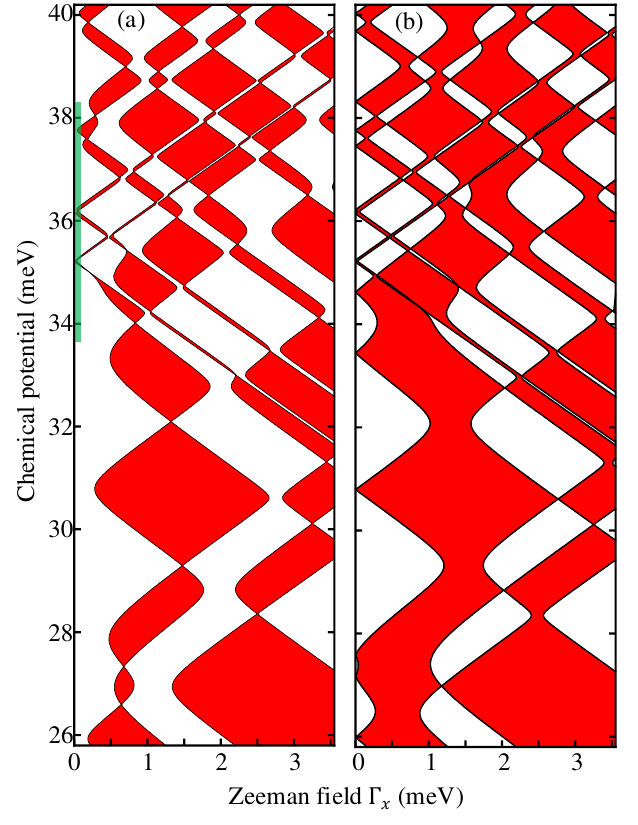}
\end{center}
\caption{Topological phase diagram as a function of Zeeman field ($\Gamma_x$) and chemical potential for a JJ structure with outside SM regions of width $W_L= W_R= 195~$nm (all other parameters being the same as in Fig. \ref{Fig9}).  The ``fragmentation'' of the topological phase diagram for $\mu\gtrsim V_{L(R)}$ is associated with the presence of $k=0$ low-energy modes having large spectral weight inside the unproximitized outside SM regions. The DOS along the zero field cut marked by the green line in (a) is shown in Fig. \ref{Fig11}.}
\label{Fig10}
\vspace{-2mm}
\end{figure}

First, we recalculate the topological phase diagrams shown in Fig. \ref{Fig3} for a system with finite outside regions and applied potential $V_L=V_R=35~$meV, comparable to the values of the chemical potential. The results corresponding to $W_L=W_R=100~$nm are shown in Fig. \ref{Fig9}, while the case $W_L=W_R=195~$nm is illustrated in Fig. \ref{Fig10}. For chemical potential values lower than the applied gate potential, i.e., when the outside SM regions are depleted, the phase boundaries are practically identical to those shown in Fig. \ref{Fig3}. By contrast, when the electrons can access the outside SM regions (i.e., for $\mu \gtrsim V_{L(R)}$), the phase diagram becomes ``fragmented'' due to the emergence of narrow features that disperse (approximately) linearly in the $\Gamma-\muup$ plane. The ``density'' of these features (hence, the ``fragmentation'' of the phase diagram) increases with increasing the width $W_{L(R)}$ of the SM regions.   

\begin{figure}[t]
\begin{center}
\includegraphics[width=0.48\textwidth]{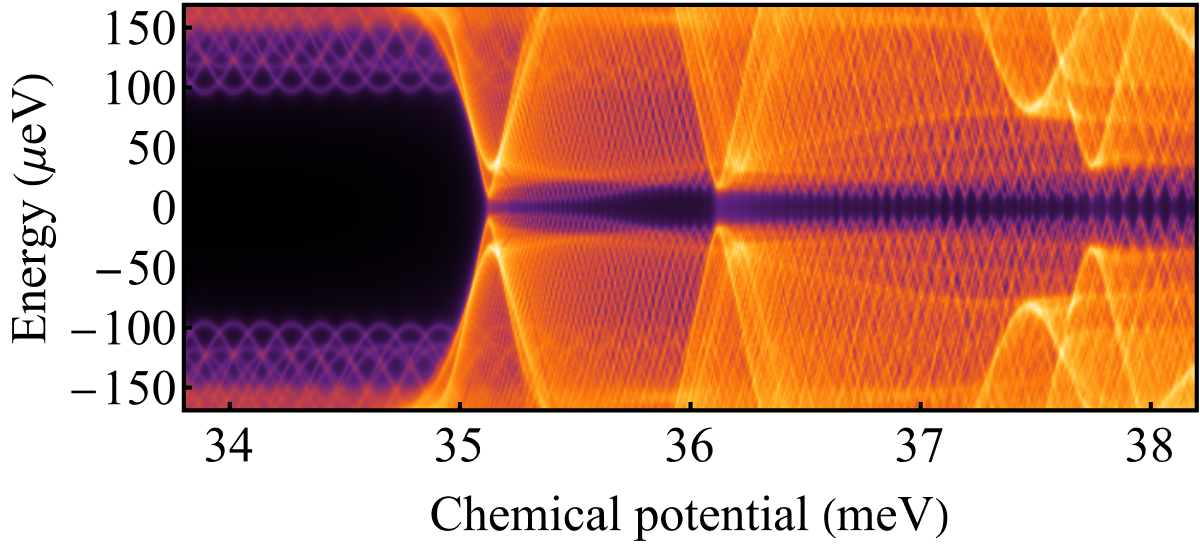}
\end{center}
\vspace{-2mm}
\caption{Dependence of the low-energy DOS on energy and chemical potential along the $\Gamma = 0$ cut marked in Fig ~\ref{Fig10}(a). Note that the quasiparticle gap collapses for $\mu\gtrsim V_{L(R)}$. Upon widening the outside SM regions, the regime  $\mu > V_{L(R)}$ becomes nearly gapless. Experimentally, one can access this regime by tuning the applied gate potential, while the chemical potential remains fixed (see Fig. \ref{Fig13}).}
\label{Fig11}
\vspace{-2mm}
\end{figure}

To clarify the physics associated with this behavior and understand its possible practical implications, we calculate the density of states as a function of chemical potential and energy along the zero-field cut marked by a green line in Fig. \ref{Fig10}(a). The results shown in Fig. \ref{Fig11} reveal a sharp ``transition'' from a large--gap regime for $\mu\lesssim V_{L(R)}$ to a small--gap regime when the outside regions become occupied with electrons, i.e., for  $\mu\gtrsim V_{L(R)}$. We point out that the system becomes effectively gapless for $\mu\gtrsim V_{L(R)}$ in the limit of wide SM regions, $W_{L(R)} >> W_J$. 
Experimentally, this zero-field ``transition'' from a gapped state to an effectively gapless state at $\mu \approx V_{L(R)}$ allows for the precise determination of the gate potential associated with the depletion of the SM regions. Assuming similar lever arms for the left, right, and junction gates (see Fig. \ref{Fig1}), this gate potential value can provide useful information about the $V_J$ values associated with the optimal regime, which typically corresponds to an almost depleted junction region (see Sec. \ref{TopoGap}). Furthermore, if one can estimate the lever arm, the ``transition'' illustrated in Fig. \ref{Fig11} enables the direct estimation of the chemical potential. 

\begin{figure}[b]
\begin{center}
\includegraphics[width=0.48\textwidth]{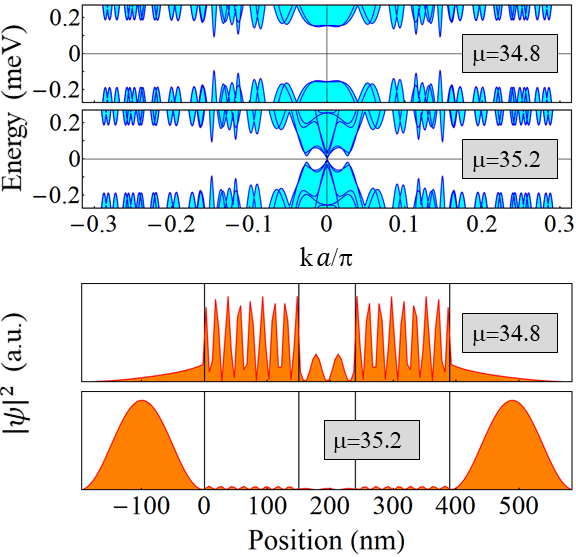}
\end{center}
\vspace{-2mm}
\caption{{\em Top}: Energy spectra for the system in Fig. \ref{Fig10}(a) at zero Zeeman field and two different values of the chemical potential (given in meV in the corresponding panels). When $\mu$ becomes (slightly) larger than $V_{L(R)}=35~$meV the quasiparticle gap collapses in the vicinity of $k=0$. {\em Bottom}: Position dependence of the lowest energy modes with $k=0$ corresponding to the spectra in the upper panels. The origin of the horizontal axis corresponds to the left edge of the left SC strip. For $\mu=34.8~$meV most of the spectral weight is distributed within the SC regions, while for $\mu=35.2~$meV the state is primarily located within the outside SM regions.}
\label{Fig12}
\vspace{-2mm}
\end{figure}

To gain further intuition regarding the physics associated with the collapse of the quasiparticle gap, we calculate the energy spectra on the two sides of the ``transition'' in Fig. \ref{Fig11}. The results shown in the upper panels of Fig. \ref{Fig12} reveal that the collapse of the gap is associated with low-$k$ modes, while the states with $k a/\pi \gtrsim 0.05$ remain practically unaffected by the small change of chemical potential. Next, we focus on the lowest energy modes with $k=0$ and calculate the spatial dependence of the corresponding probability distributions. The results shown in the bottom panels of Fig. \ref{Fig12} clearly demonstrate that the collapse of the quasiparticle gap is due to weakly proximitized states located within the outside SM regions, which emerge once these regions become accessible, i.e., for $\mu\gtrsim V_{L(R)}$. Indeed, for $\mu=34.8~$meV the state is mostly located within the SC regions and decays (roughly exponentially) within the outside SM regions, away from the SC strips. By contrast, for $\mu=35.2~$meV the state is almost entirely located within the SM regions and, consequently, has negligible induced pairing. Upon increasing the width of the SC regions, these ``normal'' modes proliferate, making the system metallic (i.e., gapless).    

\begin{figure}[t]
\begin{center}
\includegraphics[width=0.48\textwidth]{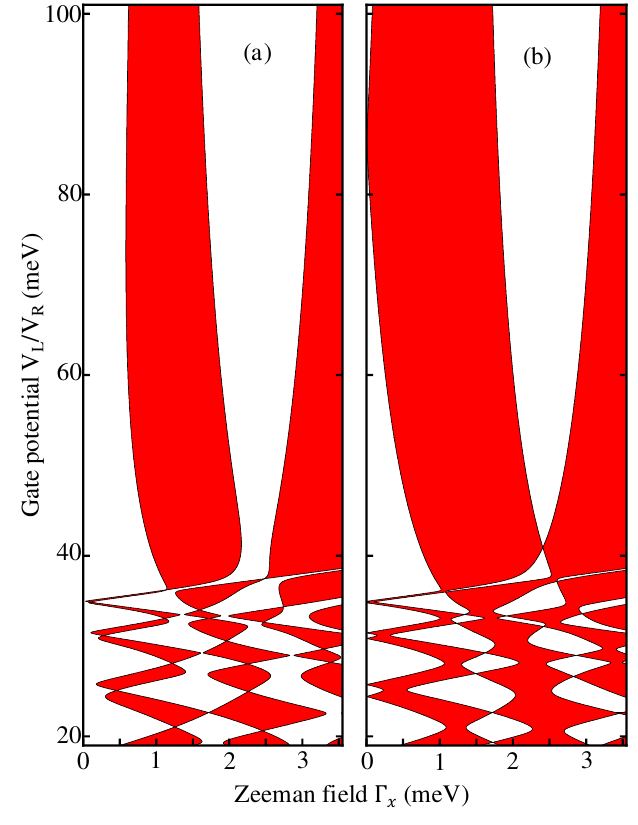}
\end{center}
\caption{Topological phase diagram as a function of Zeeman field and applied gate potentials $V_L=V_R$ for a system with $W_L= W_R= 100~$nm, chemical potential $\mu= 36~$meV, and relative phase difference between the SC films is  (a) $\phi=0$ and (b) $\phi=\pi$. Note that the (nearly gapless) regime signaled by the ``fragmentation'' of the topological phase diagram can be accessed by lowering the applied gate potentials in the outside SM regions to satisfy the condition $V_{L(R)} < \mu$.}
\label{Fig13}
\vspace{-2mm}
\end{figure}

We should emphasize that experimentally the chemical potential is not a control parameter, as it has a fixed value for a given hybrid structure. Engineering heterostructures with optimal values of the chemical potential represents a critical growth and device fabrication task, as this parameter plays an important role in realizing robust topological states \cite{Paudel2024} (also see below). In practice, the physics discussed in this section can be investigated by varying the applied gate potentials in the SM regions. As shown in Fig. \ref{Fig13}, the ``transition'' between a regime characterized by depleted outside SM regions and the regime $\mu > V_{L(R)}$ can be tuned by changing the gate potential applied to the SM regions. As expected, the $\mu > V_{L(R)}$ regime is characterized by a ``fragmented''  phase diagram, reflecting the nearly gapless nature of the underlying quantum states. In this regime, for $W_{L(R)}>>W_J$ the system becomes metallic. Finally, we point out that a reduction of the induced SC gap is generally associated with the presence of undepleted SM regions, regardless of the device geometry (e.g., in structures having SC films with ``holes''). 

\subsection{Depleted junctions and the wire-JJ crossover} \label{Crossover}

\begin{figure}[t]
\centering
\includegraphics[width=0.48\textwidth]{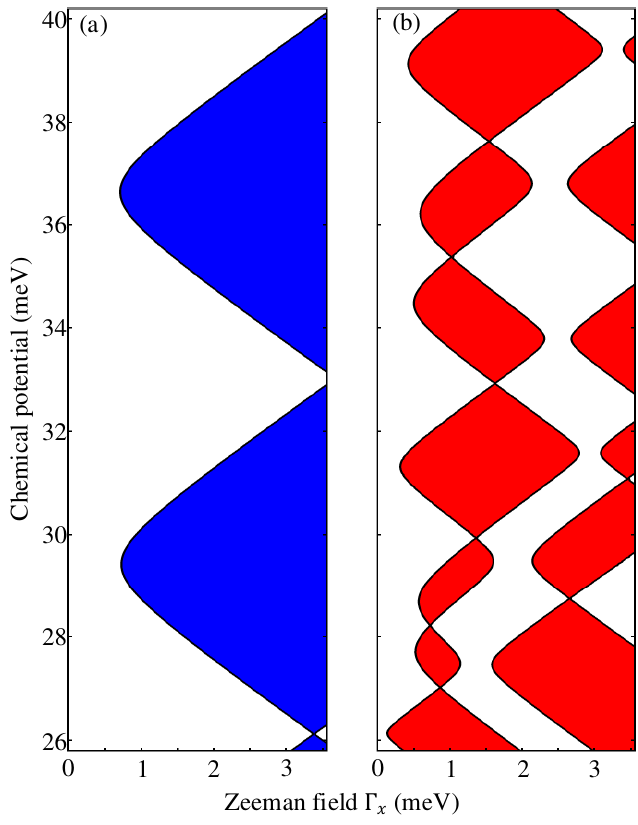}  
\caption{Comparison of topological phase diagrams obtained in (a) the ``nanowire regime'' with  $V_J = 45~$meV and (b) the ``Josephson junction regime'' with $V_J = 25~$meV. Note that the system in (a) has a depleted junction region, being equivalent to a pair of disconnected wires. The shaded regions are topologically nontrivial. The system parameters are the same as in Fig. \ref{Fig3}(a), except the $V_J$ value in panel (a).}
\label{Fig14}
\vspace{-2mm}
\end{figure}

Having investigated the effect of varying the gate potentials in the outside SM regions and the possibility of  exploiting the corresponding features for estimating experimentally the value of the chemical potential, we now turn our attention to the effects of changing the gate potential in the junction region, $V_J$. For simplicity, we assume that the outside SM regions are depleted, i.e., $V_{L(R)}> \mu$. We start with the manifest dichotomy between the regime characterized by $V_J > \mu$, when the junction region is depleted and the system consists of two decoupled hybrid nanowires, and the regime $V_J < \mu$ associated with the presence of nontrivial topological phases in planar Josephson junctions. Typical topological phase diagrams corresponding to the two regimes are shown in Fig. \ref{Fig14}. Panel (a), which corresponds to $V_J > \mu$, shows the characteristic ``hyperbolic'' topological regions (blue shading) of quasi-1D hybrid wires, with Zeeman field minima occurring at chemical potential values corresponding to the $k=0$ energies of the confinement-induced subbands. More specifically, the two large topological regions (blue shading) correspond to subbands with zero-field minima at approximately $29.4~$meV and $36.5~$meV, respectively. Note that, in this regime, there are two (identical) decoupled wires that support topological superconducting phases and associated MZMs for parameter values within the topological regions of Fig. \ref{Fig14}(a). Panel (b), on the other hand, corresponds to $V_J < \mu$ and shows a typical topological phase diagram for a planar JJ structure, which, in fact, is identical to the phase diagram in Fig. \ref{Fig3}. We emphasize that the only difference between the regimes illustrated in the two panels is the value of the applied gate potential in the junction region, $V_J$.

\begin{figure}[t]
\centering
\includegraphics[width=0.48\textwidth]{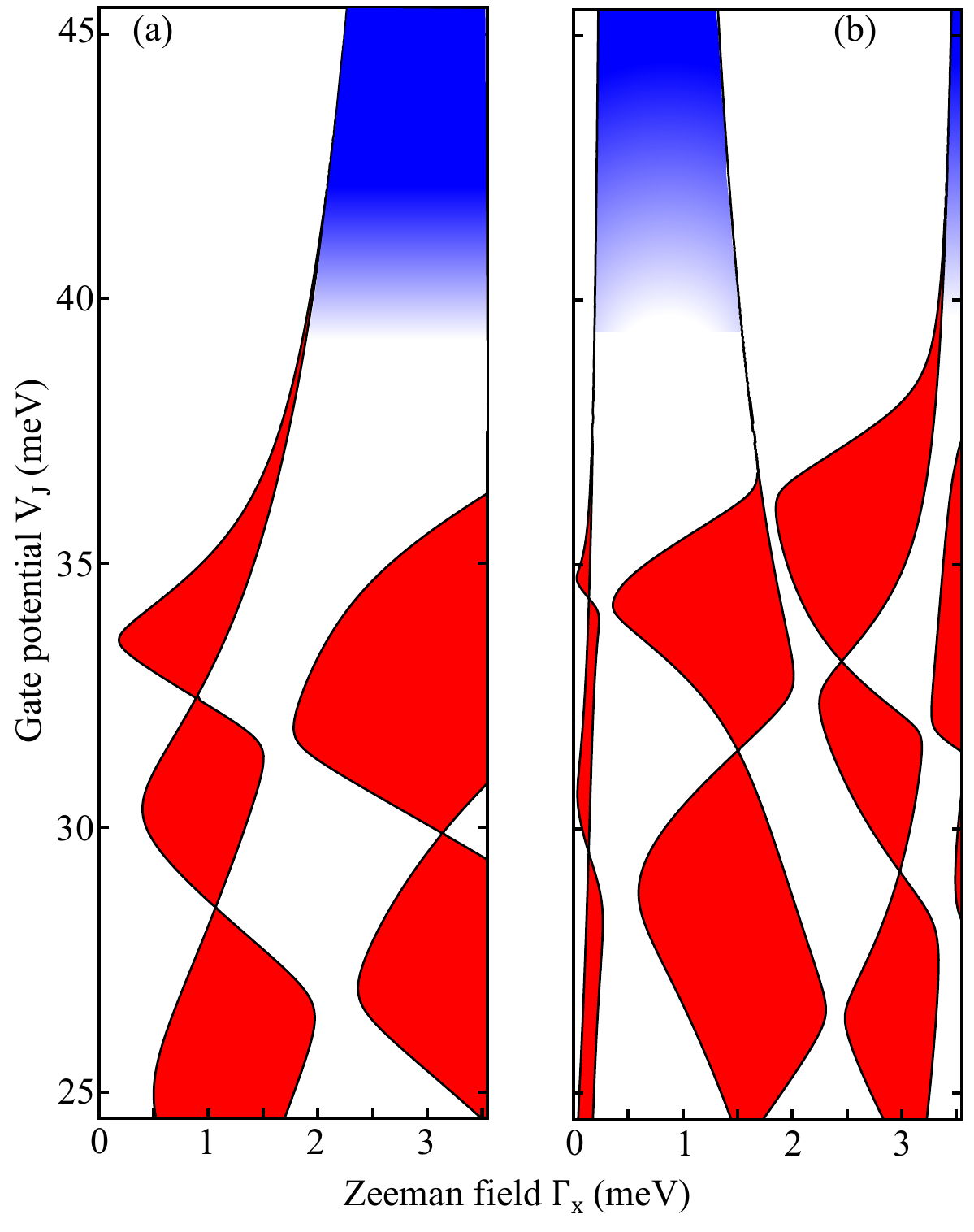}  
\caption{Crossover between the ``nanowire'' and ``Josephson junction'' regimes for a system with chemical potential $\mu=35~$meV,  $W_L = W_R = 100~$nm, and applied gate potentials in the outside SM regions (a) $V_L= V_R = 100~$meV and  (b) $V_L= V_R = 35~$meV.  Note that the ``nanowire'' topological regions (blue) are adiabatically connected to trivial ``Josephson junction'' regions, while the topological JJ phase (red) collapses into the phase boundaries associated with the ``nanowire'' regime.}
\label{Fig15}
\vspace{-2mm}
\end{figure}

A natural question arises regarding the ``transition'' from the ``nanowire'' regime to the ``Josephson junction'' regime driven by the junction potential. For concreteness, we fix the chemical potential, $\mu=35~$meV, and vary $V_J$ and the Zeeman field, all other parameters being the same as in Fig. \ref{Fig14}. The results shown in Fig. \ref{Fig15}(a) reveal a crossover between the two regimes, with the topological ``nanowire'' region (blue) being adiabatically connected to trivial ``Josephson junction'' regions and the topological JJ phase (red) collapsing into phase boundaries in the ``nanowire'' regime, i.e., upon depleting the junction region. Similar features characterize the system with almost filled outside SM regions (having $\mu=V_L=V_R$) shown in Fig. \ref{Fig15}(b). Physically, the topological ``nanowire'' regime (blue) corresponds to two decoupled nanowires, each hosting a pair of MZMs localized at the ends. Upon lowering the gate potential $V_J$, the two Majoranas at each end of the system become coupled and acquire a finite energy, rendering the corresponding SC phase topologically trivial. On the other hand, the JJ topological phase (red) requires a finite coupling between the two SC films. Increasing $V_J$ above the chemical potential (i.e., depleting the junction region) reduces this coupling, causing the JJ topological phase to collapse.

\subsection{Dependence of the topological gap on the width of the SC films} \label{TopoGap}

In the previous sections we have investigated qualitatively~ the effects of changing the~ orientation of the applied field and varying the gate potentials. We already noticed that, despite the fact that large control parameter regions are consistent with the presence of a topological superconducting phase, the (bulk) quasiparticle gaps protecting this phase are typically small and can even vanish. Therefore, a critical task for practically realizing topological superconductivity and MZMs in planar JJ structures is to identify optimal parameter regimes characterized by large values of the topological gap.  Computationally, this is a nontrivial task due, in part, to the large number of parameters characterizing the system. Here, we focus on an important component of this critical task by addressing the following question: Is there an optimal value of the width $W_{SC}$ of the superconducting films and, if yes, how does this value depend on the chemical potential? The general strategy for addressing the optimization problem is sketched in Sec. \ref{Summ}.   

\begin{figure}[t]
\centering
\includegraphics[width=0.48\textwidth]{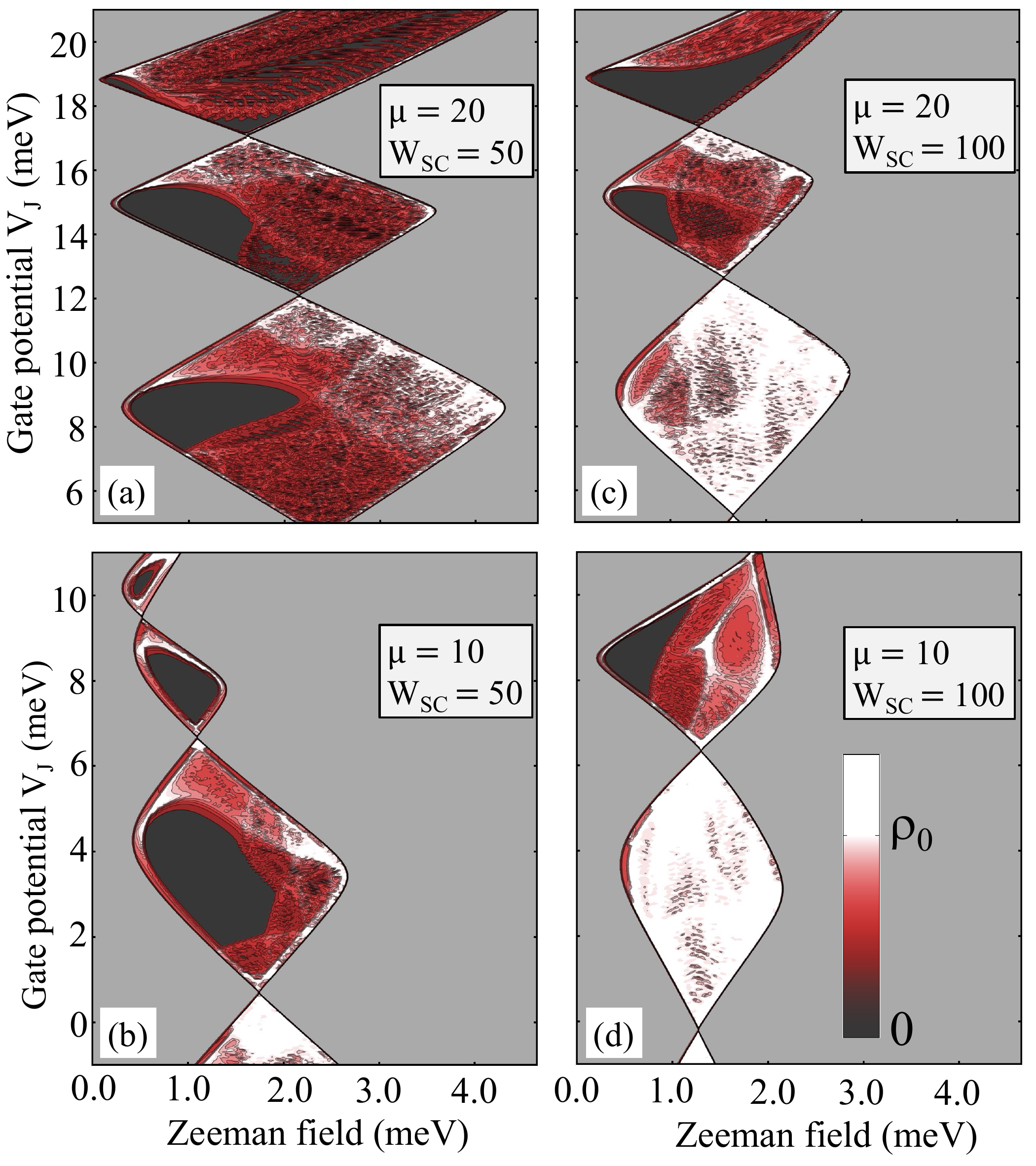}  
\caption{Comparison of the DOS at $E^*=35~\mu$eV in systems with ultra-thin SCs.
The colored areas represent the DOS within the lowest-field topological region at energy $E^*=35~\mu$eV, with black corresponding to zero DOS, i.e., gap values larger than $35~\mu$eV, and white representing DOS values larger than a reference value $\rho_0$ associated with gaps much smaller than $E^*$. 
Note that for the device with $W_{SC}=50~$nm (left panels), large topological gap regions emerge at $V_J$ values significantly smaller than the chemical potential, in contrast to the typical behavior illustrated in Fig. \ref{Fig16}. In the figure, $\mu$ is given in meV and $W_{SC}$ in nanometers.}
\label{Fig18} 
\vspace{-2mm}
\end{figure}

To answer the specific question formulated above, we consider planar JJ structures with SC film widths ranging from $W_{SC}=50~$nm to $W_{SC}=400~$nm and chemical potential values $\mu=10$, $20$, and $40~$meV. We restrict our analysis to systems with no SC phase difference, $\phi=0$, but point out that we expect similar values of the maximum gap in systems with phase difference $\phi=\pi$ (see Appendix \ref{AppA}). 
Also, we consider a fixed value of the junction width, $W_{J}=90~$nm, consistent with typical experimental values. The effect of varying $W_J$ is briefly discussed in Appendix \ref{AppD}.
The size of the topological gap can be evaluated with high accuracy using the effective Hamiltonian approximation, as discussed in the context of Fig. \ref{Fig2}. Specifically, if $E_1(k)\geq 0$ is the lowest (positive) energy mode corresponding to a set of control parameters $(\Gamma, V_J)$ within the topological region, the topological gap is $\Delta_{\rm top}(\Gamma, V_J)={\rm Min}_k[E_1(k)]$. Thus, for a system with specified values of the chemical potential ($\mu$) and SC film width ($W_{SC}$), finding the maximum topological gap and the corresponding (optimal) control parameters, $(\Gamma, V_J)$, involves diagonalizing the effective Hamiltonian ${\cal H}_{eff}(k)$ given by Eq. (\ref{Heff}) for a sufficiently dense set of $(\Gamma, V_J, k)$ points, with $(\Gamma, V_J)$ within the topological region and $k\lesssim k_F$ (where $k_F$ is a parameter-dependent Fermi wave vector). This involves a substantial computational effort. 

\begin{figure*}[t]
\centering
\includegraphics[width=0.95\linewidth]{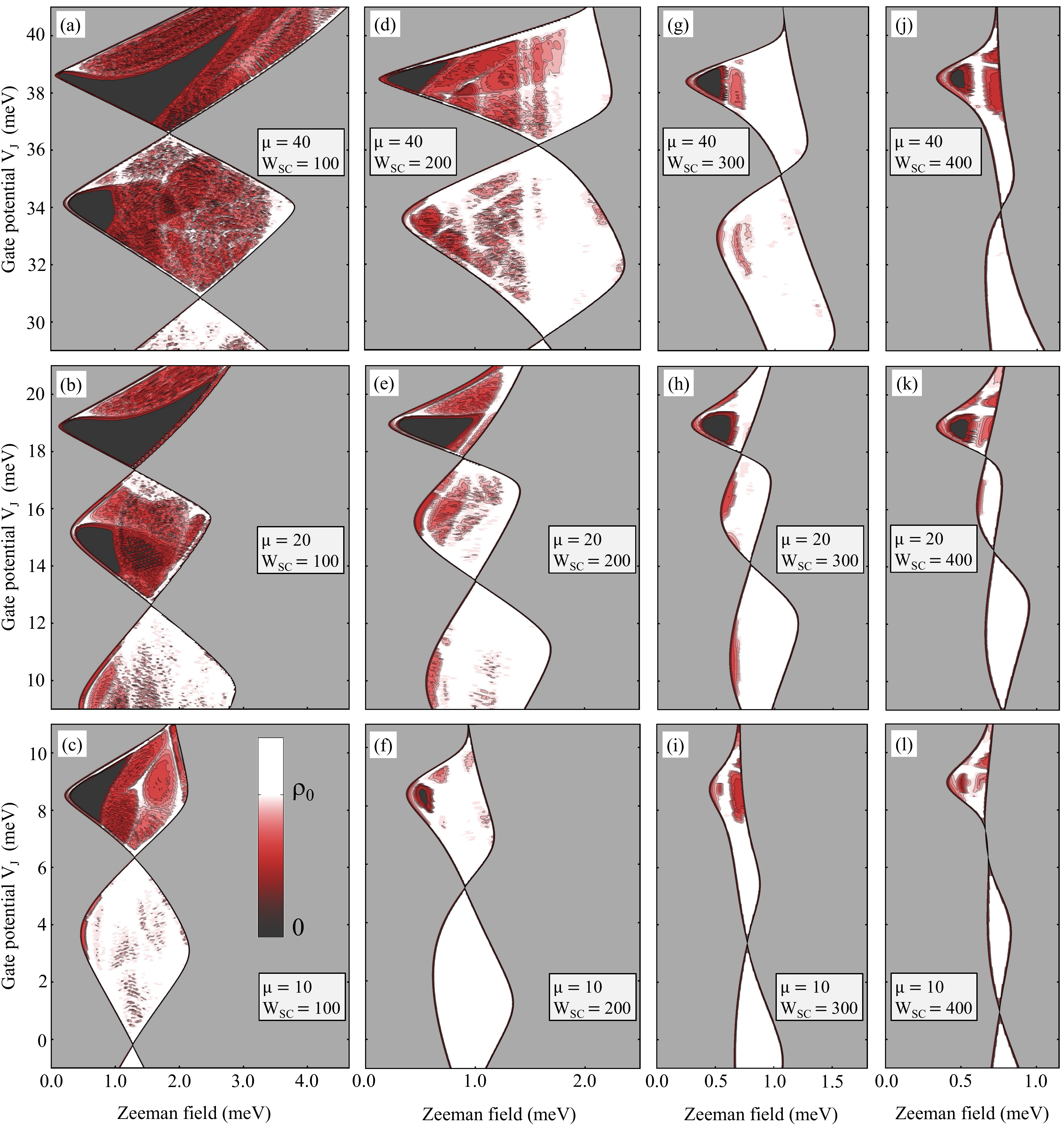}  
\caption{Survey of the large topological gap regimes for JJ devices with narrow superconductors ($W_{sc} = 100$, $200$, $300$, and $400~$nm, from left to right) and different chemical potential values ($\mu = 40$, $20$, and $10~$meV, from top to bottom). 
The colored areas represent the DOS within the lowest-field topological region at energy $E^*=35~\mu$eV, with black corresponding to zero DOS, i.e., gap values larger than $35~\muup$eV. 
Note that the large-gap regions typically correspond to a regime characterized by values of the potential $V_J$ applied in the junction region comparable to (but smaller than) the chemical potential.}
\label{Fig16} 
\vspace{-2mm}
\end{figure*}

\begin{figure*}[t]
\centering
\includegraphics[width=\linewidth]{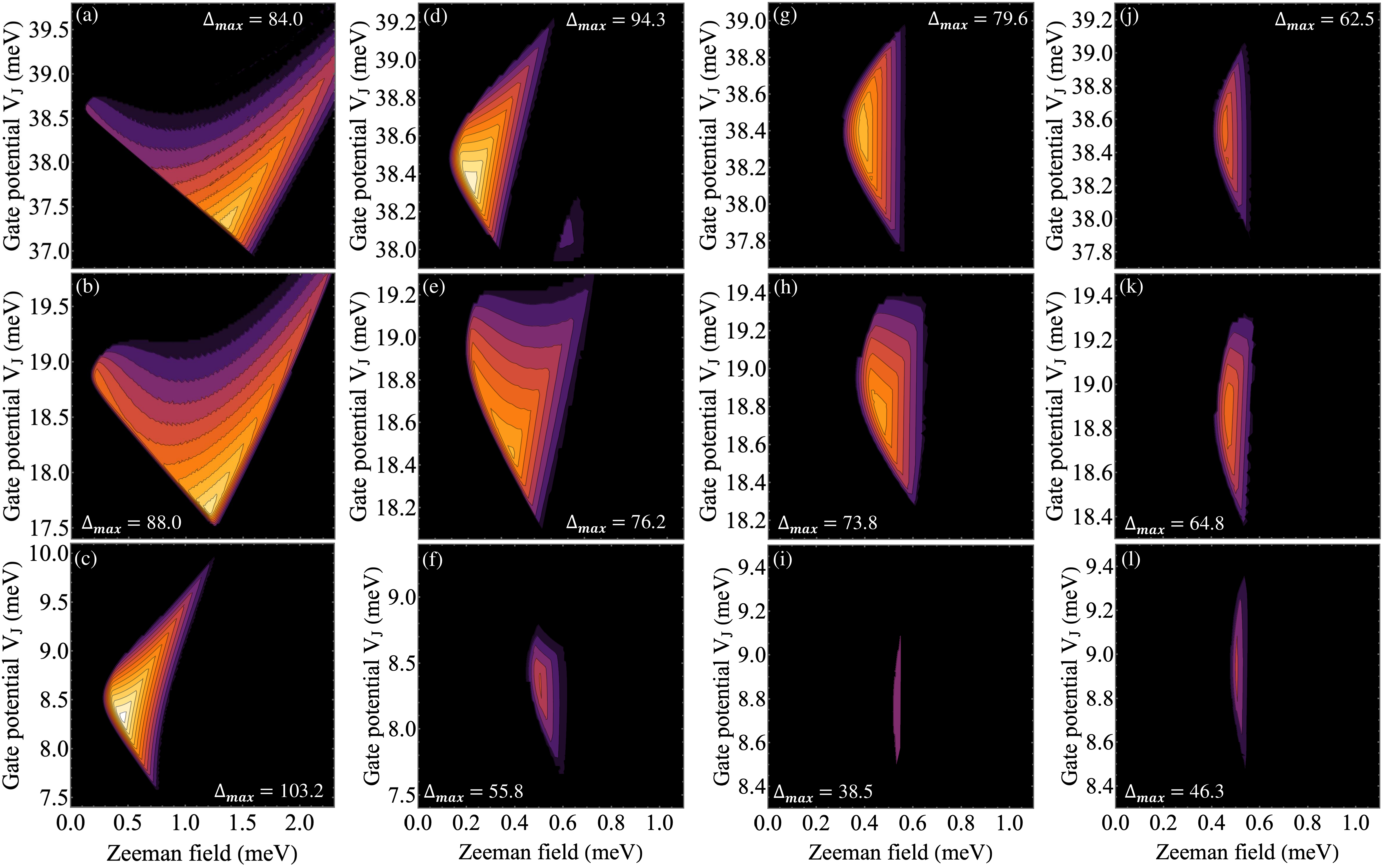}  
\caption{Topological gap within the large gap regions shown in Fig. \ref{Fig16} as a function of the gate potential $V_J$ and the applied Zeeman field. 
The rows and columns correspond to different chemical potential values  ($\mu = 40,~20,~{\rm and}~10~$meV, from top to bottom) and SC widths ($W_{sc} = 100,~200,~300,~{\rm and}~400~$nm, from left to right), as in Fig. \ref{Fig16}. The maximum values of the topological gap (in $\muup$eV) are indicated in each panel. The the contours correspond to variations of the gap size by $5~\muup$eV, with black indicating gap values smaller than $35~\muup$eV and/or topologically-trivial gaps. Note the non-monotonic dependence of the topological gap on $\mu$ and the SC width. For $\mu=40~$meV, the maximum gap is obtained for a SC width $W_{SC}\approx 200$nm. For lower chemical potential values the optimal regime corresponds to $W_{SC}\approx 100$nm (also see Figs. \ref{Fig18} and \ref{Fig19}).}
\label{Fig17}
\vspace{-2mm}
\end{figure*}

To solve the problem more efficiently, we address it using the following two-step approach. First, we calculate the density of states (DOS) within the low-field topological region at a reference energy $E^*=35~\muup$eV. We focus on the low-field regime because it is experimentally-accessible (hence, relevant). A finite DOS implies the presence of low-energy states (with energies on the order of $35~\muup$eV, or lower), while a nearly vanishing DOS signals the presence of a quasiparticle gap larger than $E^*$. The results shown in Fig. \ref{Fig18} and Fig. \ref{Fig16} reveal the following trends. For $W_{SC}\geq 200~$nm and chemical potential values up to $40~$meV, the large topological gap regime corresponds to a nearly depleted junction region. In other words, for $W_{SC}\geq 200~$nm the low DOS regions (dark shading) in Fig. \ref{Fig16} occur within the top topological ``lobe'' characterized by gate potential values $V_J$ smaller than, but comparable to the chemical potential, $\mu$. By contrast, for ultra-thin SC films (Fig. \ref{Fig18} and left panels in Fig. \ref{Fig16}) large gap regions also occur within lower topological ``lobes'', which give the dominant contributions for $W_{SC} = 50~$nm (see Fig. \ref{Fig18}).  
In addition, we note that the area of the large gap regions increases with decreasing the width $W_{SC}$ of the SC films and with increasing the chemical potential. 

\begin{figure}[t]
\centering
\includegraphics[width=0.48\textwidth]{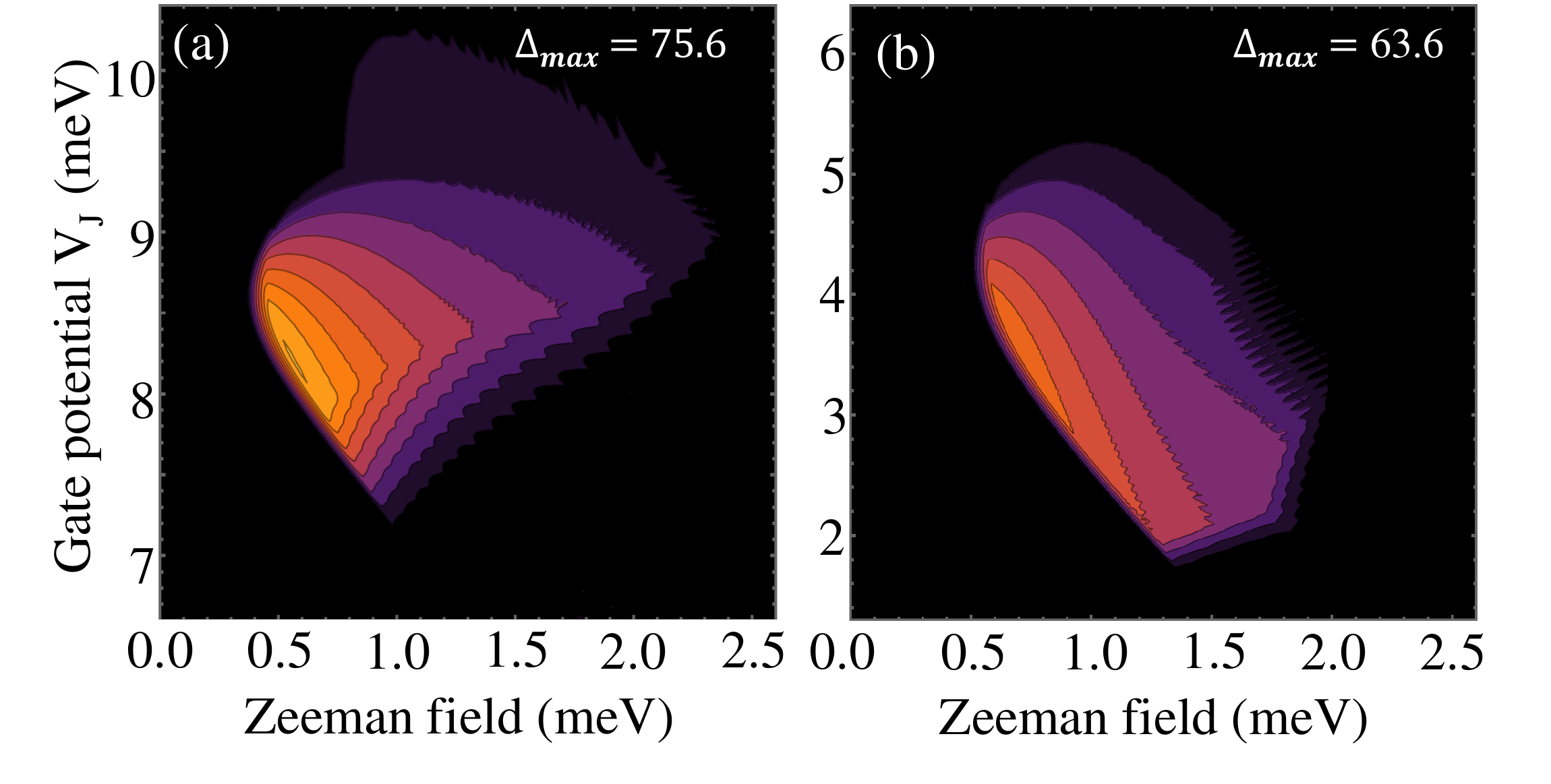}  
\caption{Topological gap for a device with $W_{SC}=50~$nm and chemical potential (a) $\mu=20~$meV and (b) $\mu=10~$meV. Note that the other large gap regions in Fig. \ref{Fig17}(a) and (b) are characterized by smaller values of the maximum topological gap. Comparison with the results in Fig. \ref{Fig16} corresponding to $W_{SC}=100~$nm indicates that the optimal SC width for $\mu\lesssim 20~$meV is about $100~$nm.}
\label{Fig19}
\vspace{-2mm}
\end{figure}

The second step of our approach involves calculating the topological gap (only) within the large gap regions identified in the first step. By construction, this implies calculating the topological gap for all control parameter values consistent with $\Delta_{\rm top} > 35~\muup$eV. The results shown in Figs. \ref{Fig17} and \ref{Fig19} provide our answer to the question posed at the beginning of this section. Specifically, we find that, for a given value of the chemical potential, the maximum topological gap depends non-monotonically on the width of the SC films, with an optimal width value $W_{SC}^*(\mu)$ that increases with increasing the chemical potential.  The ``optimal width'', $W_{SC}^*(\mu)$, is defined as the SC width corresponding to the largest value of $\Delta_{\rm max}(W_{SC},\mu)$, for a given chemical potential. Thus, based on the results in Fig. \ref{Fig17} and Fig. \ref{Fig19}, we find that $W_{SC}^*(10)\sim 100~$nm, $W_{SC}^*(20)\sim 100~$nm, and 
$W_{SC}^*(40)\sim 200~$nm, where the $\mu$ values are given in meV. Upon further increasing $\mu$, we expect larger values of $W_{SC}^*$, but still within the range of a few hundred nanometers. Note that our analysis is intended to capture the significant trends, not the detailed, quasi-continuous dependence of the maximum topological gap on the SC width and the chemical potential, which is likely to exhibit small local fluctuations in the $W_{SC}-\mu$ plane. The low-energy modes that control the topological gap are discussed in App. \ref{AppB}. In essence, these are low-k modes associated with the top few occupied subbands (see App. \ref{AppB}). Thus, the size of their characteristic quasiparticle gaps depends strongly of the spin-orbit coupling strength. A brief discussion of the role played by the spin-orbit coupling and of the effect of varying the parent superconducting gap is provided in App. \ref{AppC}. In addition, the topological gap depends (non-monotonically) on the width, $W_J$, of the junction region. As shown in Appendix \ref{AppD}, for JJ structures with narrow superconductors of given width, $W_{SC}$, the optimal values of the junction width that maximize the topological gap are comparable to the width of the SC, $W_J\sim W_{SC}$.  

The key conclusion of our analysis is that the optimal SC width is within the $100-200~$nm range and increases (slightly) with increasing the chemical potential. Furthermore, the maximum topological gap corresponding to the optimal regime has values on the order of $90-100~\muup$eV, which represents about $35-40\%$ of the parent SC gap ($\Delta_0=250~\muup$eV). This is a significant value, quite larger than the typical topological gap values (up to $\approx 25\%$ of $\Delta_0$) characterizing planar JJ structures with wide SC films (other parameters being similar) \cite{Paudel2021,Paudel2024} and comparable to the largest topological gaps emerging in similar (clean) SM-SC hybrid nanowires \cite{Paudel2024}. Considering that planar JJ structures are likely to be less susceptible to disorder than nanowires realized using 2D structures with similar parameters \cite{Paudel2024}, optimizing the SC width opens a promising route toward realizing robust topological superconducting phases and Majorana zero modes.

\section{Summary and discussion} \label{Summ}

A critical task associated with the realization of robust topological quantum devices is to solve the corresponding optimization problem and identify the parameter regimes that maximize the resilience of the emerging topological quantum states (e.g., against disorder). In this paper, we have addressed a key aspect of the optimization problem in the context of semiconductor-superconductor planar Josephson junctions predicted to host topological superconductivity and Majorana zero modes by focusing on the dependence of the topological gap on the width of the superconducting films for different values of the chemical potential, while continuously tuning the control parameters (i.e., the Zeeman field and the gate potential in the junction region). We have considered ``ideal'' (i.e., clean and infinitely long) systems modeled using an effective Green's function approach, as well as an effective Hamiltonian formulation. The Green's function of the 2D semiconductor [see Eq. (\ref{Gwk})] is calculated based on a tight-binding Hamiltonian [Eq. (\ref{HSM})], with a contribution induced by the proximity coupling to the superconducting films incorporated as a self-energy term [Eq. (\ref{SigSC})]. The effective Bogoliubov--de Gennes Hamiltonian, on the other hand, incorporates the proximity-induced effects through both a pairing term and a renormalization factor [Eq. (\ref{Heff})]. The two methods are equivalent at zero energy and in excellent agreement at low energies, i.e., below $\sim 100~\muup$eV. 

We first discuss the generic features of the topological phase diagram in planar Josephson junctions with narrow superconductors and show that, upon rotating the Zeeman field, the regions in parameter space characterized by nontrivial values of the $\mathbb{Z}_2$ topological invariant remain finite, but the system may become gapless (Sec. \ref{General}). Next, we consider the effect of varying the gate potential applied within the semiconductor regions outside the superconducting strips and find that  the regime characterized by electrostatic potential values lower than the chemical potential, when these regions become occupied with electrons, is associated with a collapse of the induced pairing potential at zero magnetic field (Sec. \ref{SSB}). On the other hand, tuning the gate potential applied within the junction region drives a crossover between a Majorana nanowire regime corresponding to a depleted junction region and a Josephson junction regime corresponding to gate potential values lower than the chemical potential (Sec. \ref{Crossover}). Finally, using the insights provided by this preparatory analysis, we address the optimization problem by considering systems having different chemical potential values and superconducting film widths (Sec. \ref{TopoGap}). To deal with the computational challenges, we adopt a two-step approach, first calculating the density of states (DOS) within the low-field topological region at an energy $E^*=35~\mu$eV, then determining the topological gap (only) within the low-DOS regions. We find that the topological gap is maximized in structures with superconductor films of width ranging between $100~$nm, and $200~$nm, comparable to the width of the junction region ($W_J=90~$nm). The optimal superconductor width increases slightly with the chemical potential. The corresponding maximum values of the topological gap can be as high as $40\%$ of the parent superconducting gap, significantly larger than topological gap values obtained in wide-superconductor structures with comparable materials and control parameters.    

The present study is a component of a general, multi-step strategy that we propose for addressing the optimization problem and, ultimately realizing robust planar Josephson junction topological devices. 
{Step I}: Establish a theoretical base line by identifying the optimal regimes of parameters that can be easily controlled experimentally (e.g., Zeeman field, gate potentials, and geometric parameters), while making ``reasonable'' assumptions regarding the hard-to-control/measure parameters (e.g., spin-orbit coupling and SM-SC coupling) and considering ``ideal'' conditions (e.g., no disorder and no finite size effects).  
{Step II}: Estimate the actual parameter values (including the disorder strength) that characterize the system in laboratory conditions based on a theory--experiment feedback loop. This can be done by fabricating devices based on the theoretical guidelines obtained in Step I, measuring them over a large control parameter space, and systematically comparing the results with theoretical predictions based on ``realistic'' models that incorporate disorder, finite size effects, and other experimentally-relevant factors. 
{Step III}: Identify realistic optimal conditions within the range of accessible system parameters (and/or within the close ``vicinity'' of this range), then realize and operate devices within these optimal regimes. This may require considering spatially-modulated hybrid structures \cite{Pan2021}, fine-tuning the SM-SC interface, lowering the chemical potential, or reducing the effective disorder strength. The ultimate goal is to fabricate robust devices that would enable the unambiguous demonstration of Majorana zero modes. 

While investigating in detail the impact of varying additional parameters, such as the SM-SC coupling, the size of the parent SC gap, the spin-orbit coupling strength, the phase difference between the SC films, and the width of the junction region, will be considered in Step II, we provide hints regarding the expected (qualitative) behavior in Appendices \ref{AppA}, \ref{AppC}, and \ref{AppD}. Thus,  for JJ structures with SC phase difference $\phi=\pi$ we expect the emergence of an expanded large-gap topological region (as compared to their $\phi=0$ counterparts), but comparable values of the maximum topological gap (see App. \ref{AppA}). Also, varying the size of the parent SC gap does not affect the phase boundaries and results in a nonlinear variation of the topological gap (App. \ref{AppC}). On the other hand, varying the spin-orbit coupling strength affects both the phase boundaries and the size of the topological gap, again, nonlinearly (App. \ref{AppC}). In addition, we find that the dependence of the (maximum) topological gap value on the width of the topological region is non-monotonic, with an optimal regime corresponding to junctions widths comparable to the width of the SC ribbons (App. \ref{AppD}). Regarding the expected effect of varying the effective SM-SC coupling strength, we point out that a strongly coupled JJ device is more robust against disorder and less susceptible to finite size effects than a weakly coupled device \cite{Paudel2024}. Therefore, the strongly coupled regime investigated in this work is the likely optimal regime. Nonetheless, the specific quantitative aspects associated with this issue require further investigations. Finally, we emphasize that the most significant factor that is not considered in this work is disorder, which affects both the superconductor and the semiconductor heterostructure. Estimating the relevant disorder types and corresponding strengths is a crucial component of Step II.

We conclude with a few remarks regarding our proposed strategy and the significance of the present work in this context. First, we point out that the optimal regimes characterizing clean structures may ``shift'' (in parameter space) in the presence of disorder. Nonetheless, we expect the narrow-superconductor optimal regimes identified in this study to be more robust against disorder than generic topological phases emerging in wide-superconductor devices, which are characterized by relatively small topological gaps or are gapless. Realizing devices in this ``ideal'' optimal regime is crucial for Step II, as they are likely to be closer to the ``real'' optimal regime. On the other hand, estimating the system parameters (in Step II) is a key component of our plan, as a brute force numerical optimization over a huge ``realistic'' parameter space (which includes materials and interface parameters, as well as geometric, disorder, and control parameters) is a practical impossibility. Finally, we point out that, once the parameter values characterizing the system in laboratory conditions are determined, the optimization must also account for finite size effects that may impact the stability of the topological superconducting phase and the corresponding Majorana modes. In particular, this may impose constraints on the acceptable values of the chemical potential, since large $\mu$ values result in the emergence of low-energy states with large characteristic length scales, i.e., strong finite size effects.

\begin{acknowledgements}
P. P. acknowledges the support from ONR N00014-23-1-2061 and from the Air Force Office of Scientific Research by the Department of Defense under Award No. FA9550-23-1-0498 of the DEPSCoR program. T.S. was supported by ONR N00014-23-1-2061, and J.S. acknowledges support from ONR N00014-22-1-2764. 
\end{acknowledgements}


\appendix
\renewcommand{\thefigure}{A\arabic{figure}}
\setcounter{figure}{0}

\section{Topological gap in planar JJ structures with superconducting phase difference $\phi=\pi$}\label{AppA}

In Sec. \ref{TopoGap}, we focus on the optimization problem for JJ structures with no superconducting phase difference, $\phi=0$. A natural question concerns the dependence of the topological gap on the width of the SC films in the presence of a phase difference, in particular in systems with $\phi=\pi$. While this problem is beyond the scope of this study, we mention a few relevant points. First, structures with $\phi=\pi$ are characterized by ``large gap'' regions significantly bigger than their $\phi=0$ counterparts. On the one hand, this implies that solving (numerically) the optimization problem for $\phi=\pi$ involves a (substantially) larger computational cost, because the corresponding parameter space region is larger. On the other hand, identifying experimentally a large gap region provides more information in a system with no phase difference, for which the region is relatively well localized in the $\Gamma-V_J$ plane. In other words, identifying experimentally a large gap region in a system with $\phi=0$ provides (indirect) information about the chemical potential and the effective g-factor. Therefore, within our general optimization strategy investigating the $\phi=0$ case is part of Step I, while the study of JJ structures with $\phi=\pi$ is a component of Step II. 

\begin{figure}[t]
\centering
\includegraphics[width=0.43\textwidth]{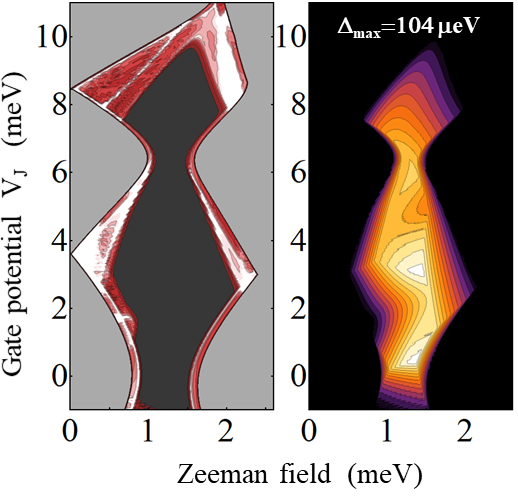}  
\caption{{\em Left}: Density of states within the lowest-field topological region at energy $E^*=35~\muup$eV for a system with $\mu=10~$meV, $W_{SC}=100~$nm, and superconducting phase difference $\phi=\pi$. Note the expanded ``large gap'' region (black), as compared to the corresponding region characterizing a JJ structure with no phase difference [see Fig. \ref{Fig16}(c)]. {\em Right}: Topological gap within the ``large gap'' region. The contours correspond to variations of the gap size by $5~\muup$eV, with the white regions having gap values above $100~\muup$eV.}
\label{FigA1}
\vspace{-2mm}
\end{figure}

Second, we point out the interesting observation that the maximum values of the topological gap corresponding to $\phi=0$ and $\phi=\pi$ were found to be comparable in several JJ structures (with different system parameters, e.g., SC widths, SM-SC couplings, etc.) studied in Refs. \cite{Paudel2021} (for straight junctions) and \cite{Paudel2024}. To strengthen this observation, we apply the procedure described in Sec. \ref{TopoGap} to a system with $\mu=10~$meV, $W_{SC}=100~$nm, and $\phi=\pi$. The results are shown in Fig. \ref{FigA1}. The left panel confirms our statement regarding the expanded large gap region, as compared to the $\phi=0$ case [for comparison see Fig. \ref{Fig16}(c)]. The corresponding topological gap map, which is shown in the right panel of \ref{FigA1}, is characterized by two regions with gap values exceeding $100~\muup$eV. However, the maximum topological gap is $\Delta_{\rm max}\approx 104~\muup$eV, less than $1\%$ larger than the maximum gap of the $\phi=0$ system [see Fig. \ref{Fig17}(c)]. This is remarkable, particularly considering that the topological gap maxima correspond to different control parameter regimes, i.e., different values of the Zeeman field, $\Gamma$, and applied junction potential, $V_J$. These observations suggest that the maximum topological gap characterizing JJ structures with $\phi=\pi$ is likely to have a non-monotonic dependence on $W_{SC}$ qualitatively similar to that identified in systems with no phase difference. Finally, we point out that, although the topological phase extends down to $\Gamma=0$ (for specific $V_J$ values), the optimal regime of a JJ structure with $\phi=\pi$ corresponds to finite (relatively large) Zeeman field values [$\Gamma \sim 1.1-1.5~$meV in Fig. \ref{FigA1} (right), versus $\Gamma \sim 0.4-0.6~$meV in Fig. \ref{Fig17}(c)]. This suggests that the ``theoretical'' advantage of having $\phi$ as an additional knob that ``shifts'' the topological phase down to $\Gamma=0$ may not play a significant role in actually realizing a robust topological phase in JJ devices. Nonetheless, having this additional knob could prove instrumental in studying the system and estimating various parameters.

\renewcommand{\thefigure}{B\arabic{figure}}
\setcounter{figure}{0}

\section{Low-energy modes and representative quantum states}\label{AppB}

\begin{figure}[t]
\begin{center}
\includegraphics[width=0.48\textwidth]{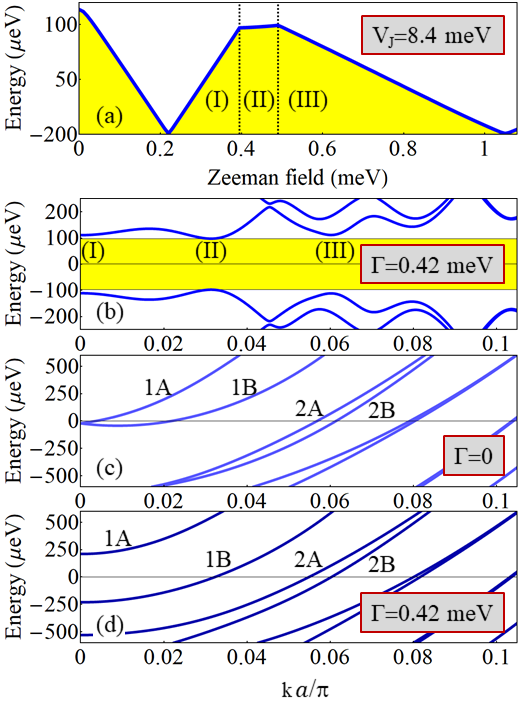}
\end{center}
\vspace{-4mm}
\caption{(a) Dependence of the quasiparticle gap on the Zeeman field for a system with SC width $W_{SC} = 100~$nm, chemical potential $\mu=10~$meV, and electrostatic potential in the junction region $V_J=8.4~$meV. Note that this corresponds to a constant $V_J$ cut slightly above the maximum topological gap in Fig. \ref{Fig17}(c), which obtains for $V_J\approx 8.3~$meV. The topological gap edge within the regions labeled (I), (II), and (III) is controlled by the corresponding low-$k$ modes shown in panel (b), which illustrates the k-dependence of the superconducting spectrum for $\Gamma=0.42~$meV. 
Note that the modes characterized by larger Fermi wave vectors (i.e., $k_F \gtrsim 0.08 \pi/a$) have larger values of the quasiparticle gap. (c) and (d) Normal phase spectra corresponding to Zeeman field values $\Gamma=0$ and $\Gamma=0.42~$meV, respectively. The zero of the energy is chosen at the Fermi level. Note that the top occupied subbands (1A, 1B, 2B, and 2B), which control the size of the topological gap, are strongly affected by the Zeeman field, in contrast to the large-$k_F$ subbands.}
\label{FigB1}
\vspace{-2mm}
\end{figure}

In this Appendix, we investigate the low-energy modes that control the size of the topological gap. First, we consider a JJ structure with SC width $W_{SC}=100~$nm and chemical potential $\mu=10~$meV and calculate the quasiparticle gap along a constant $V_J$ cut slightly above the maximum topological
gap shown in Fig. \ref{Fig17}(c). The results given in Fig. \ref{FigB1}(a) clearly indicate the presence of three different regions characterized by (I) a topological gap that increases (nearly) linearly with the Zeeman field, (II) a topological gap that depends weakly on $\Gamma$, and (III) a topological gap that decreases approximately linearly with the Zeeman field. 

\begin{figure}[t]
\begin{center}
\includegraphics[width=0.48\textwidth]{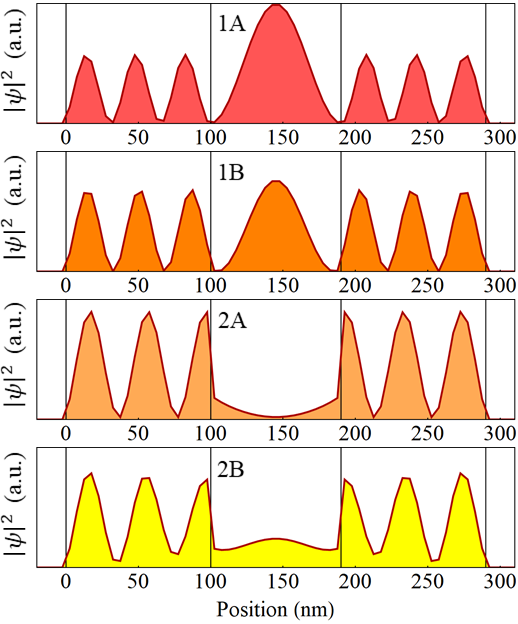}
\end{center}
\vspace{-2mm}.
\caption{Transverse profiles of the low-energy modes labeled 1A--2B in Fig. \ref{FigB1}. The system has chemical potential $\mu=10~$meV, SC width $W_{SC}=100~$nm, and junction potential $V_J=8.4~$meV. Note that the left SC film covers the region $0-100~$nm, the junction is between $100~$nm and $190~$nm, etc. Note that the spectral weight within the proximitized regions is distributed between the SM ($33\%$) and the parent SC ($67\%$) (also, see Fig. \ref{FigB6}].}
\label{FigB2}
\vspace{-2mm}
\end{figure}

We identify the low-energy modes responsible for this behavior by calculating the dependence of the (superconducting) energy spectrum on the wave vector $k$. An example, corresponding to a Zeeman field value $\Gamma =0.42~$meV, is shown in Fig. \ref{FigB1}(b). One can clearly notice three low-k modes with quasiparticle gaps at or near the topological gap edge  (in this case, $\Delta_{\rm top}\approx 98~\muup$eV), while the modes characterized by larger $k$ values have larger quasiparticle gaps. Note that $\Gamma =0.42~$meV is within region (II) [see Fig. \ref{FigB1}(a)] and the corresponding topological gap is controlled by the mode with characteristic wave vector $k\approx 0.033\pi/a$, which is marked as mode (II) in panel (b), while the modes (I) and (III) have slightly larger gaps. In region (I), the topological gap is controlled by the mode with $k=0$, while in region (III) it is controlled by the mode with characteristic wave vector $k\approx 0.06\pi/a$. Note that the energy of mode (I) vanishes for $\Gamma \approx 0.23~$meV, which corresponds to the vanishing of the $k=0$ bulk gap at the topological quantum phase transition (TQPT). The gap characterizing mode (III) becomes very small (but finite) near $\Gamma \approx 1.05~$meV. Of course, this does not indicate the presence of a TQPT [also see the map in Fig. \ref{Fig17}(c)]. 

The normal state low-energy spectrum corresponding to the superconducting spectrum in Fig. \ref{FigB1}(b) is shown in panel (d), while the normal spectrum in the absence of an applied Zeeman field is given in Fig. \ref{FigB1}(c). Clearly, the topological gap is controlled by the top occupied subbands, specifically those labeled 1A, 1B, 2A, and 2B.  Note the significant effect of the applied Zeeman field on the subbands 1A and 1B at energies near the Fermi level. This is the result of the combined effect of having weak spin-orbit coupling (because of the small values of the characteristic wave vector) and relatively large spectral weights within the junction region, where the $g$-factor is not renormalized by the parent superconductor. Indeed, the transverse profiles of the low-energy modes shown in Fig. \ref{FigB2} reveal the presence of large maxima withing the junction region for modes 1A and 1B (top panels). By contrast, the modes 2A and 2B are characterized by small amplitudes within the junction region, most of the spectral weight being within the proximitized regions [see Figs. \ref{FigB2}(c) and (d)]. This feature, combined with the larger values of the characteristic $k$-vectors associated with these modes (as compared to 1A and 1B), explains their weaker dependence on the Zeeman field in the vicinity of the Fermi level, where the relevant physics takes place. Furthermore, as shown in Fig. \ref{FigB1} [panels (c) and (d)], the impact of the Zeeman field on the modes with larger values of the characteristic wave vector is even weaker. In turn, this results in larger values of the corresponding quasiparticle gaps [see Fig. \ref{FigB1}(b)], which, upon increasing $k_F$, become comparable to the zero field values ($\sim 200~\muup$eV).

\begin{figure}[t]
\begin{center}
\includegraphics[width=0.48\textwidth]{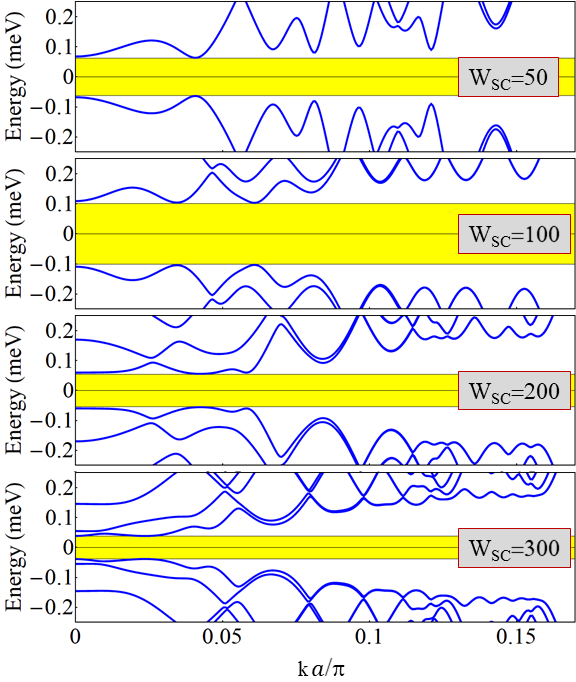}
\end{center}
\caption{Dependence of the low energy spectrum on the wave vector $k$ for JJ devices with chemical potential $\mu=10~$meV, different SC widths, $W_{SC}$ (given in nanometers), and control parameters near the maximum of the topological gap: (a) $W_{SC}=50~$nm, $V_J=3.64~$meV, $\Gamma_x=0.68~$meV; 
(b) $W_{SC}=100~$nm, $V_J=8.3~$meV, $\Gamma_x=0.45~$meV; (c) $W_{SC}=200~$nm, $V_J=8.34~$meV, $\Gamma_x=0.51~$meV; (d) $W_{SC}=300~$nm, $V_J=8.94~$meV, $\Gamma_x=0.54~$meV. Note the non-monotonic dependence of the gap on the SC width, with a maximum around $W_{SC}\approx 100~$nm.}
\label{FigB3}
\vspace{-2mm}
\end{figure}

In summary, our analysis of the low-energy modes that control the topological gap (within the large gap regime) shows that the key modes are associated with the top occupied subbands. More specifically, in region (I) the size of the topological gap is determined by a linear combination of modes 1A (dominant) and 1B (some contribution) with $k\approx 0$. In region (II), the gap is controlled by the mode 1B (with finite $k$), while in region (III) the topological gap is practically determined by the mode 2B. Note that the mode 2A is associated with slightly larger values of the quasiparticle gap. In addition, deeper occupied subbands, which are characterized by larger values of the Fermi wave vector, are associated with minima of the quasiparticle gap larger than the topological gap and, practically, they do not affect its value. 

\begin{figure}[t]
\begin{center}
\includegraphics[width=0.48\textwidth]{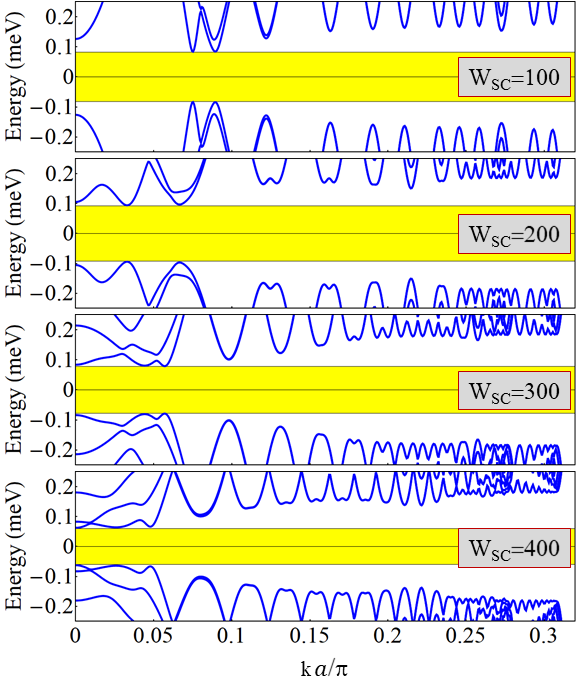}
\end{center}
\caption{Same as in Fig. \ref{FigB3} for a system with chemical potential $\mu=40~$meV and: (a) $W_{SC}=100~$nm, $V_J=37.3~$meV, $\Gamma_x=1.35~$meV; (b) $W_{SC}=200~$nm, $V_J=38.35~$meV, $\Gamma_x=0.32~$meV; (c) $W_{SC}=300~$nm, $V_J=38.4~$meV, $\Gamma_x=0.39~$meV; (d) $W_{SC}=400~$nm, $V_J=38.6~$meV, $\Gamma_x=0.45~$meV. The maximum gap corresponds to $W_{SC}\approx 200~$nm. Note that the gap size is determined by the low-$k$ modes, i.e., the modes corresponding to the top 3-4 occupied transverse bands. The lower energy occupied bands (i.e., the large $k$ modes) are characterized by significantly larger superconducting gaps.}
\label{FigB4}
\vspace{-2mm}
\end{figure}

\begin{figure}[t]
\begin{center}
\includegraphics[width=0.48\textwidth]{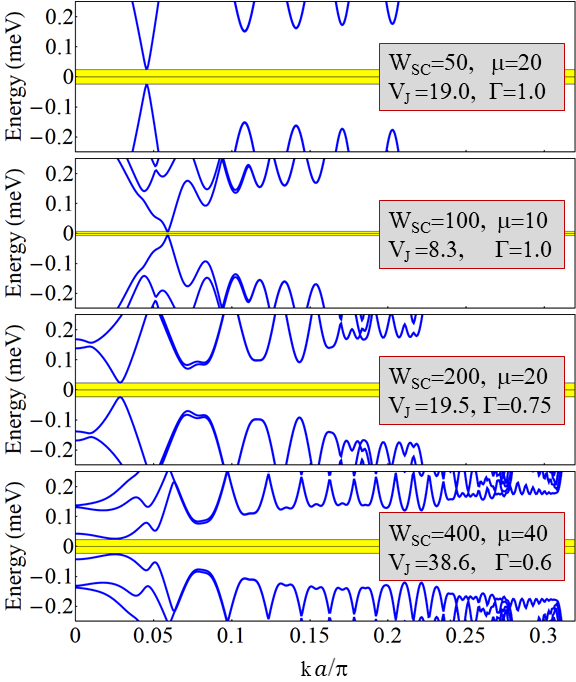}
\end{center}
\caption{Low-energy spectra corresponding to small values of the topological gap. The relevant parameters are given inside each panel ($W_{SC}$ in nm and $\mu$, $V_J$, and $\Gamma=\Gamma_x$ in meV). Note that the gap size is determined by the low-$k$ modes, i.e., the top occupied transverse bands.}
\label{FigB5}
\vspace{-2mm}
\end{figure}

This picture is further strengthened by the low-energy superconducting spectra shown in Fig. \ref{FigB3} and Fig. \ref{FigB4}, which correspond to systems with chemical potential values $\mu=10~$meV and $\mu=40~$meV, respectively, different SC widths, and control parameters near the maxima of the topological gap. Indeed, in all cases illustrated in the two figures the topological gap (highlighted by the yellow shading) is controlled by a few top occupied subbands with low characteristic wave vectors up to about $0.1\pi/a$. By contrast, the deep occupied bands are associated with larger values of the quasiparticle gap. The notable exception corresponding to the top panel of Fig. \ref{FigB3} is due to the fact that in systems with ultra-thin SCs (e.g., $W_{SC}=50~$nm) the maximum topological gap does not emerge within the top ``topological lobe'' (see Fig. \ref{Fig18}).  In addition, the low-gap regions within the top ``topological lobes'' (see Fig. \ref{Fig16}) are also controlled by modes associated with the top occupied subbands, as illustrated by the examples shown in Fig. \ref{FigB5}. Thus, the low-energy physics of the top ``topological lobes'' (as well as the physics of nearby trivial regions) is completely controlled by the properties of a small number (3-4) of top occupied subbands, the deeper occupied bands being associated with larger gap values. 

We point out that for the system with $\mu=10~$ meV and $W_{SC}=100~$nm the maximum topological gap emerges at $V_J\approx 8.3~$meV, when the gap minima associated with modes 1A, 1B, and 2B have practically equal values (see Fig. \ref{FigB3}) and region (II) (see Fig. \ref{FigB1}) has shrunk to a point. Lowering $V_J$ reduces regions (I) and (III) and, implicitly the maximum gap, while increasing $V_J$ generating a finite region (II) with an increasing width and decreasing maximum gap values. In systems with larger chemical potential values the large gap regions (see Fig. \ref{Fig17}) are dominated by wide type-(II) regions, i.e., by the 1B mode with finite $k$. Lowering $V_J$ reduces the width of this region and enhanced the gap, but the corresponding topological phase ``migrates'' toward larger values of the Zeeman field (see Figs. \ref{Fig16} and \ref{Fig17}). The net result is that, although the (large) topological gap of all systems with $W_{SC}=100~$nm is controlled by the same low-energy modes (in particular 1A, 1B and 2B), increasing the chemical potential corresponds to satisfying the maximum condition at larger values of the Zeeman field, which result in decreasing the $\Delta_{max}$ values (also see Fig. \ref{Fig17}).

\begin{figure}[t]
\begin{center}
\includegraphics[width=0.48\textwidth]{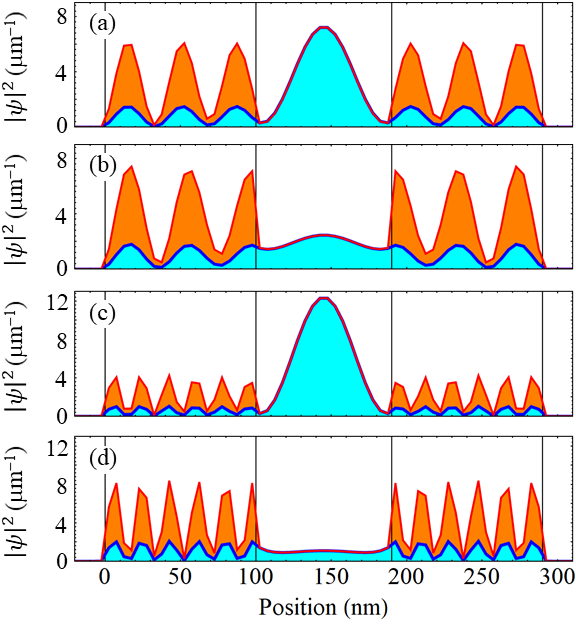}
\end{center}
\caption{Position dependence of the gap-edge states corresponding to the system with $W_{SC}=100~$nm and $\mu=10~$meV in Fig. \ref{FigB3} (top two panels) and $W_{SC}=100~$nm and $\mu=40~$meV in Fig. \ref{FigB4} (bottom two panels). The corresponding values of the wave vector are: (a) $k=0.034\pi/a$, (b) $k=0.061\pi/a$, (c) $k=0.075\pi/a$, (d) $k=0.089\pi/a$. The red line (with orange shading) represents the wave function amplitude, $|\psi|^2$, calculated using the effective Hamiltonian method, while the blue line (with cyan shading) represents the local density of states calculated using the Green's function approach. Note that in the (central) junction region the two quantities coincide, while in the proximitized regions the LDOS corresponds to $Z~\!|\psi|^2$, with $Z=(1+\gamma/\Delta_0)^{-1}=1/4$, since only a fraction $Z$ of the spectral weight is located inside the SM (the other being located inside the parent SC). Asso note that (a) and (b) correspond to the modes 1B and 2B in Fig. \ref{FigB2}.} 
\label{FigB6}
\vspace{-2mm}
\end{figure}

Finally, we note that the (transverse) spatial profiles of the relevant low-energy modes associated with the large gap regimes depend on the chemical potential. As an example, in Fig. \ref{FigB6} we compare the profiles of the 1B and 2B modes associated with the (finite-$k$) gap minima corresponding to the maximum topological gap of the system with $W_{SC}=100~$nm and chemical potential values as in Figs. \ref{FigB3} and \ref{FigB4}. The top panels correspond to a system with $\mu=10~$mev, while the bottom panels are for a JJ structure with $\mu=40~$meV. In Fig. \ref{FigB6}, we also provide a comparison of the position dependence of $|\psi|^2$ calculated using the effective Hamiltonian approach and the position dependence of the corresponding (properly normalized) local density of states (LDOS). The two profiles practically concide within the junction region, while in the proximitized region the LDOS is reduced by a factor of $4$, as it contains only the spectral weight located within the semiconductor ($3/4$ of the local spectral weight being within the parent superconductor). 

\renewcommand{\thefigure}{C\arabic{figure}}
\setcounter{figure}{0}

\section{Dependence on the parent superconducting gap and the spin-orbit coupling}\label{AppC}

In this Appendix, we briefly consider the dependence of the topological gap (within the large gap regime of control parameters) on the size of the parent superconducting gap and the strength of the spin-orbit coupling. As discussed in the main text, the full investigation of the dependence on these parameters (as well as on the SM-SC strength and  disorder parameters) is beyond the scope of this study, being part of Step II of our proposed strategy. For concreteness, we focus on a system with SC width $W_{SC}=100~$nm and chemical potential $\mu=10~$meV. 

\begin{figure}[t]
\begin{center}
\includegraphics[width=0.48\textwidth]{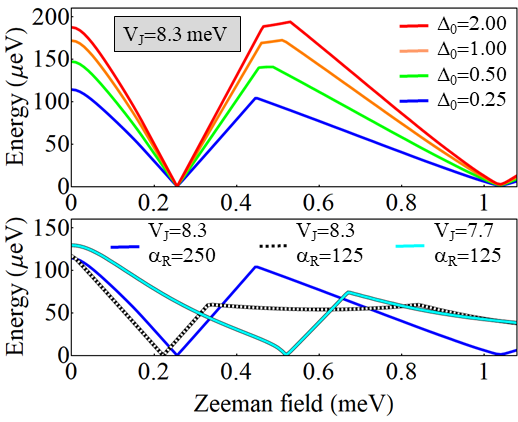}
\end{center}
\caption{{\em Top}: Dependence of the quasiparticle gap on the Zeeman field for a JJ structure with SC width $W_{SC}=100~$nm, chemical potential $\mu=10~$meV, junction potential $V_J=8.3~$meV, and different values of the parent SC gap $\Delta_0$ (given in meV). The vanishing of the gap at $\Gamma_c\approx 0.26~$meV, which signals a TQPT, is independent of $\Delta_0$. {\em Bottom}: Dependence of the quasiparticle gap on the Zeeman field for a JJ structure with SC width $W_{SC}=100~$nm, and chemical potential $\mu=10~$meV. The values of the junction potential $V_J$ are given in meV, while the Rashba spin-orbit coupling $\alpha_R=\alpha\cdot a$ is given in meV$\cdot$\AA . Increasing $\Delta_0$ by a factor of eight roughly doubles the maximum topological gap, while reducing the spin-orbit coupling strength by a factor of two results in a decrease of the maximum topological gap by about $30\%$. } 
\label{FigC1}
\vspace{-2mm}
\end{figure}

In the top panel of Fig. \ref{FigC1} we show the dependence of the quasiparticle gap on the applied Zeeman field for fixed junction potential ($V_J=8.3~$meV) and different values of the parent superconducting gap, $\Delta_0$. First, we point out that the TQPT (at the critical field $\Gamma_c\approx 0.26~$meV) is independent of the parent SC gap. This property is a result of the self-energy $\Sigma_{SC}(\omega)$ given by Eq. (\ref{SigSC}) becoming independent of $\Delta_0$ at zero frequency, $\omega=0$. We note that the minimum near $\Gamma\approx 1.05~$meV corresponds to a smal but finite gap and is not associated with a TQPT. 

Second, we note that, as expected, the topological and trivial gaps increase with increasing $\Delta_0$, but the enhancement reduces as the parent gap becomes larger. The value of the applied junction potential, $V_J=8.3~$meV, corresponding to the optimal topological gap for $\Delta_0=0.25~$meV (blue line), generates a finite type--(II) region for a system with $\Delta_0>0.25~$meV (see Fig. \ref{FigB1} and the corresponding text in App. \ref{AppB}). Therefore, the maximum values of the topological gap for $\Delta_0>0.25~$meV, which are obtained at (slightly) lower values of the applied potential, $V_J$, are (slightly) larger than the values of the gap shown in Fig. \ref{FigC1}. Nonetheless, a rough estimate indicates that doubling the size of the original gap (i.e., having $\Delta_0=0.5~$meV) corresponds to an enhancement by about $50\%$ of the (maximum) topological gap, while increasing the parent SC gap eight times (i.e., $\Delta_0=2~$meV) roughly doubles the size of the maximum topological gap. Further increasing $\Delta_0$ generates a relatively small enhancement of the topological gap.

This somewhat limited possibility of enhancing the topological gap through increasing the size of the parent superconducting gap is due to the dependence of the topological gap on other system parameters, in particular the spin-orbit coupling strength. Note that the emergence of a finite region (II) in the top panel of Fig. \ref{FigC1} (for $\Delta_0 > 0.25~$meV) already indicates that the gap is controlled by the 1B mode (see App. \ref{AppB}), which is characterized by a quasiparticle gap strongly dependent on the (effective) spin-orbit coupling. To further emphasize this point, we consider the (optimal) system with $\Delta_0=0.25~$meV and reduce the Rashba coupling strength by a factor of two, to $\alpha_{_R} =\alpha\cdot a =125~$meV$\cdot$\AA. The corresponding dependence of the quasiparticle gap on the Zeeman field is given by the dashed line in the bottom panel of Fig. \ref{FigC1}. Note the large type-(II) region characterized by a gap value representing about $55-60\%$ of the original maximum gap ($\Delta_{max}\approx 103~$meV; blue curve). The maximum gap for the system with $\alpha_{_R} =125~$meV$\cdot$\AA~ is $\Delta_{max} \approx 74~\mu$eV (i.e., about $72\%$ of the ``original'' maximum gap) and is obtained for $V_J\approx 7.7~$meV (cyan curve). 

This analysis emphasizes the importance of identifying the ``weak link'' that characterizes JJ structures realized experimentally, i.e., the parameter that imposes the most restrictive constraint on the size of the topological gap. Optimizing the other parameters without addressing the ``weak link'' problem can generate a small-to-moderate enhancement of the topological gap, while mitigating the ``weak link'' constraint can lead to a substantial gap enhancement. In addition, one has to explicitly incorporate the effects of disorder. Apriori, it is not obvious that a larger topological gap automatically implies a more stable topological phase. For example, upon increasing the ratio between the effective SM-SC coupling and the parent SC gap, $\gamma/\Delta_0$, the system becomes less susceptible to disorder inside the SM and more susceptible to SC disorder. Hence the optimal regime in the presence of disorder is generally different from that of a clean system and depends on the specific types of disorder present in the system and on the corresponding disorder parameters. 

\renewcommand{\thefigure}{D\arabic{figure}}
\setcounter{figure}{0}

\section{Dependence on the width of the junction region}\label{AppD}

In this Appendix, we briefly consider the dependence of the maximum value of the topological gap  on the width of the topological region, $W_J$. The full investigation of the dependence on this parameter is part of Step II of our proposed strategy. The key point that we want to convey is that, for a given width of the superconducting leads, $W_{SC}$, the maximum topological gap (within the lower field topological region) depends non-monotonically on the width $W_J$ of the junction region, with an optimal regime corresponding to $W_J$ values comparable to the width of the (narrow) SC strips.

\begin{figure}[t]
\begin{center}
\includegraphics[width=0.48\textwidth]{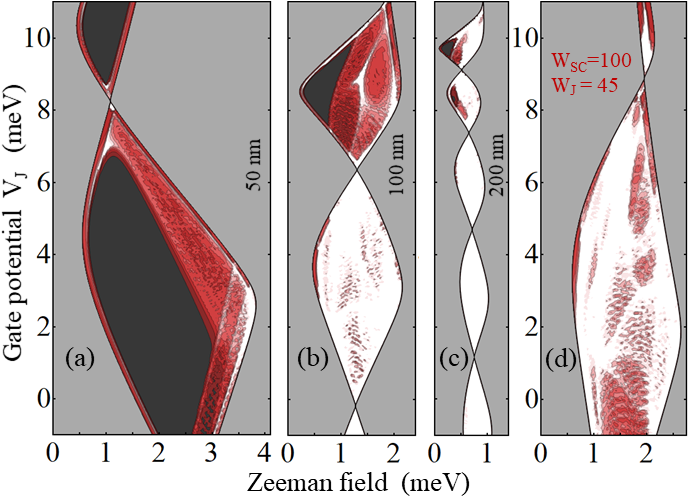}
\end{center}
\caption{Comparison of the DOS at $E^*=35~\muup$eV for systems with junction and SC regions having different widths. The systems in (a), (b), and (c) are characterized by the same ratio $W_{J}/W_{SC}=0.9$, with $W_{SC}$ values given explicitly in the corresponding panels and $W_J=45,~90,~{\rm and}~180~$nm, respectively. Panel (d) shows the DOS (at $E^*=35~\muup$eV) for a system with $W_{SC}=100~$nm and an ultra-narrow junction, $W_J=45~$nm.}
\label{FigD1}
\vspace{-2mm}
\end{figure}

For concreteness, we focus on a system with chemical potential $\mu=10~$meV and different values of $W_J$ and $W_{SC}$ corresponding to a given ratio $W_J/W_{SC}=0.9$. The results shown in panels (a), (b) and (c) of Fig. \ref{FigD1} represent the density of states (DOS) at $E^*=35~\muup$eV as function of Zeeman field (within the low-field region) and applied gate potential $V_J$ for a system with (a) $W_{SC}=50~$nm ($W_{J}=45~$nm), (b) $W_{SC}=100~$nm ($W_{J}=90~$nm) --- same as in Fig. \ref{Fig16}(d) and Fig. \ref{Fig17}(c), and (c) $W_{SC}=200~$nm ($W_{J}=180~$nm). In addition, in Fig. \ref{FigD1}(d) we present the DOS at  $E^*=35~\muup$eV for a system with $W_{SC}=100~$nm and $W_{J}=45~$nm.

\begin{figure}[t]
\begin{center}
\includegraphics[width=0.48\textwidth]{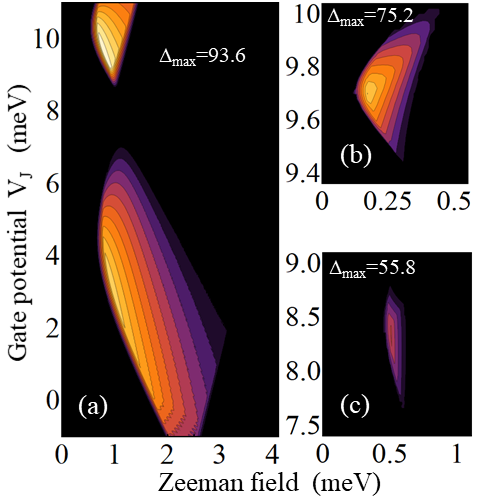}
\end{center}
\caption{(a) Topological gap as a function of Zeeman field and chemical potential for a JJ structure with ultra-thin SC strips, $W_{SC}=50~$nm, junction width $W_J=0.45~$nm (i.e., $W_{J}/W_{SC}=0.9$), and chemical potential $\mu=10~$meV, i.e., the same parameters as in Fig. 28(a). Note that the maximum topological gap ($\Delta_{max}=93.6~\muup$eV) in this device is significantly larger than the maximum topological gap in Fig. 19(b), which corresponds to a wider junction, $W_{J}/W_{SC}=1.8$, all other parameters being the same. (b) Topological gap within the ``large gap region'' in Fig. 28(c) corresponding to a JJ structure with $W_{SC}=200~$nm, $W_J=180~$nm (i.e., $W_{J}/W_{SC}=0.9$), and $\mu=10~$meV. The maximum topological gap value supported by  this structure is larger than the maximum gap of the corresponding to a device with a narrower junction region, $W_{J}/W_{SC}=0.45$ (all other parameters being the same), i.e., larger than the maximum topological gap value in Fig. 18 (f), which is shown again here [panel (c)], for convenience.}     
\label{FigD2}
\vspace{-2mm}
\end{figure}

First, we notice that the number of ``lobes'' associated with the low-field topological region increases with $W_J$, from two in panels (a) and (d) ($W_j=45~$nm) to five in panel (c) ($W_j=180~$nm), reflecting a weakening of the confinement of the relevant modes within the junction region. Second, a direct comparison of panels (a), (b), and (c) shows that the large gap regions are not uniquely determined by the ratio $W_J/W_{SC}$, but depend on both $W_J$ and $W_{SC}$. However, for a given SC width, the topological gap is maximized for values of the junction width comparable to $W_{SC}$, i.e., ratios $W_J/W_{SC}$ close to one.

To strengthen this point, we explicitly calculate the size of the topological gap within the large gap regions and compare the maximum values corresponding to systems with a given SC width and different $W_J$ values. Thus, for a system with ultra-narrow SC trips ($W_{SC}=50~$nm), decreasing the junction width from $W_J=90~$nm (i.e., $W_J/W_{SC}=1.8$) --- see Figs. \ref{Fig18}(b) and \ref{Fig19}(b) --- to $W_J=45~$nm (i.e., $W_J/W_{SC}=0.9$) --- see Figs. \ref{FigD1}(a) and \ref{FigD2}(a) --- results in an enhancement of the maximum topological gap from $\Delta_{max}=63.6~\muup$eV to $\Delta_{max}=93.6~\muup$eV. In other words, junction widths much larger than $W_{SC}$ are suboptimal. On the other hand, having very narrow junctions (relative to the SC width) is also detrimental. For example, a system with $W_{SC}=200~$nm [see Figs. \ref{Fig17}(f), \ref{Fig16}(f), \ref{FigD1}(c), and \ref{FigD2}(b) and (c)] has a maximum gap $\Delta_{max}=75.2~\muup$eV for $W_J/W_{SC}=0.9$ [Fig. \ref{FigD2}(b)], as compared to $\Delta_{max}=55.8~\muup$eV for $W_J/W_{SC}=0.45$ [Fig. \ref{FigD2}(c)]. Similarly, the large gap $\Delta_{max}=103.2~\muup$eV characterizing the $W_{SC}=100~$nm system with $W_J/W_{SC}=0.9$ [Fig. \ref{Fig18}(c)] gets reduced substantially down to $\Delta_{max}=33.2~\muup$eV upon narrowing the junction width so that $W_J/W_{SC}=0.45$ [Fig. \ref{FigD1}(d)].

These examples clearly illustrate our point that, in systems with narrow superconducting strips, the optimal topological regime corresponds to $W_J$ values comparable to the SC width. WE emphasize, however, that a detailed quantitative analysis is beyond the scope of this study, as it involves taking into account the possible impact of other factors (e.g., varying the strengths of the effective SM-SC coupling and spin-orbit coupling and including disorder effects). For example, while in clean structures the topological gap is maximized for system widths on the order of a few hundred nanometers, the presence of disorder may shift the optimal topological regime toward wider structures, to fully leverage the advantage of two-dimensionality over the one-dimension constraint in mitigating disorder effects (see, e.g., Ref. \cite{Paudel2024}).

\bibliography{References}
\end{document}